\documentclass[aps,prd,12pt,citeautoscript,superscriptaddress,showpacs,scrartcl]{revtex4-1}
\usepackage[utf8x]{inputenc}

\usepackage{xcolor}
\usepackage[colorlinks=true, pdfstartview=FitV,linkcolor=blue, citecolor=blue, urlcolor=blue]{hyperref}  

\usepackage[bottom]{footmisc}

\usepackage{graphicx}
\usepackage{upgreek}

\makeatletter
\newcommand*\bigcdot{\mathpalette\bigcdot@{.5}}
\newcommand*\bigcdot@[2]{\mathbin{\vcenter{\hbox{\scalebox{#2}{$\m@th#1\bullet$}}}}}
\makeatother

\newcommand{\be}{\begin{eqnarray}}
\newcommand{\ee}{\end{eqnarray}}
\newcommand{\bea}{\begin{eqnarray}}
\newcommand{\eea}{\end{eqnarray}}

\newcommand{\nn}{\nonumber}

\newcommand{\blue}{\textcolor{black}}
\newcommand{\purple}{\textcolor{black}}

\usepackage{natbib}
\usepackage{graphicx}
\usepackage{amsmath}
\usepackage{amssymb}
\usepackage{mathtools}
\usepackage{tensor}
\usepackage{accents}
\usepackage{cancel}
\usepackage{titlesec}
\usepackage{subfigure}
\usepackage[a4paper]{geometry}
\usepackage{psfrag}
\usepackage{latexsym,array,theorem,mathrsfs,subfigure,bm,float}
\usepackage{amsfonts}
\usepackage{amssymb}

\newcommand{\chio}{\chi}
\newcommand{\chioo}{\omega}

\newcommand{\m}{{\cal M}}

\renewcommand{\theequation}{\arabic{section}.\arabic{equation}}

\newcommand{\A}{{\rm a}}
\newcommand{\R}{{\rm r}}
\newcommand{\T}{{\rm t}}
\newcommand{\J}{{\rm J}}
\newcommand{\M}{{\rm M}}
\newcommand{\MU}{{\upmu}}

\newcommand{\nuu}{{a^2}}

\newcommand{\V}{{V}}

\newcommand{\K}{{K}}
\newcommand{\W}{W}

\newcommand{\C}{\eta}

\newcommand{\rK}{\Sigma}

\DeclareMathOperator{\diag}{diag}
\DeclareMathOperator{\e}{e}

\titleformat{\paragraph}
{\normalfont\normalsize\bffseries}{\theparagraph}{1em}{}
\titlespacing*{\paragraph}
{0pt}{3.25ex plus 1ex minus .2ex}{1.5ex plus .2ex}

\begin{document}

\title{Stationary generalizations  for  the vacuum ring wormhole}

\author{Mikhail~S.~Volkov}
\email{mvolkov@univ-tours.fr}
\affiliation{
Institut Denis Poisson, UMR - CNRS 7013, \\ Universit\'{e} de Tours, Parc de Grandmont, 37200 Tours, France}

\begin{abstract} 

\vspace{1 mm}

The ring wormhole is the zero-mass limit of the Kerr metric. Its
geometry is locally flat, but the topology is nontrivial,
with a throat connecting two asymptotic regions and a distributional curvature singularity on
the ring encircling the throat.
We construct stationary generalizations of this static wormhole that are different from Kerr and
invariant under reflections across the wormhole throat.
The problem
reduces to solving the vacuum Ernst equations subject to the corresponding symmetry conditions.
The slowly rotating perturbative solutions
were constructed previously, 
while we now present a detailed analysis of non-perturbative solutions obtained within a
numerical framework. 
For slow rotation, they exhibit the non-relativistic relation
$M\sim J^2$ between the mass and angular momentum,
which transforms into the
Regge relation $J=M^2$ in the fast-rotation regime, when $J\to\infty$ and the
ring is stretched without bound by the centrifugal force.
However, if the ring size in the static limit is sent to zero at the same time, then $M$ and $J$ remain bounded
as the throat linear velocity approaches unity.
The wormhole geometry then approaches the extremal Kerr solution, thus ``mimicking'' it.
The wormholes carry a curvature singularity at the ring, but this can be removed by
via simple ``scalarization'' procedure that promotes the vacuum solutions to regular wormholes with
a phantom scalar field.

\end{abstract} 

\maketitle

\section{Introduction}

Wormholes are bridges or tunnels between
different universes or different parts of the same universe. They were first introduced
by Einstein and Rosen (ER) \cite{Einstein:1935tc}, who noticed that the Schwarzschild black hole
actually possesses two exterior regions connected by a bridge. The ER bridge is spacelike
and cannot be traversed by classical objects, but it
may perhaps connect quantum particles, thereby producing quantum entanglement
and the Einstein-Pololsky-Rosen (EPR) effect \cite{Einstein:1935rr}, hence ER=EPR \cite{Maldacena:2013xja}
(see \cite{Javed:2025hvr} for a recent discussion).
Wormholes were also considered as geometric models of elementary particles -- handles of space
trapping an electric flux inside, for example -- a description that
may indeed be valid at the Planck scale \cite{Misner:1957mt}. Wormholes can also describe initial data
for the Einstein equations \cite {Misner:1960zz} (see \cite{Cvetic:2014vsa} for a recent review),
whose time evolution
corresponds to the observed black hole collisions \cite{Abbott:2016blz}.

An interesting topic is traversable wormholes, which are accessible to ordinary classical particles
or light \cite{Morris:1988tu} (see \cite{Visser:1995cc} for a review).
In the simplest case, such
a wormhole is described by a static, spherically symmetric line element
\be \label{oo1}
ds^2=-A^2(x)dt^2+dx^2+r^2(x)(d\vartheta^2+\sin^2\vartheta d\varphi^2),
\ee
where $A(x)$ and $r(x)$ are symmetric under $x\to -x$ and
attain non-zero global
minima at $x=0$. If both $A$ and $r/x$ approach unity as $x\to\pm\infty$, then
the metric describes two asymptotically flat regions connected by a throat of
radius $r(0)$. The Einstein equations $G^\mu_\nu=T^\mu_\nu$
imply that the energy density $\rho=-T^0_0$ and the
radial pressure $p=T^r_r$ satisfy, at $x=0$,
\be \label{2}
\rho+p=-2\frac{r^{\prime\prime}}{r}<0,~~~~~
p=-\frac{1}{r^2}<0.
\ee
It follows that, for a static wormhole to be a solution of the Einstein equations,
the Null Energy Condition (NEC),
$T_{\mu\nu}v^\mu v^\nu=R_{\mu\nu} v^\mu v^\nu \geq 0$ for any null $v^\mu$,
must be violated.

\iffalse
Another  demonstration \cite{Morris:1988tu}  of the violation of the NEC uses the  
Raychaudhuri equation \cite{Hawking:1973uf} for a bundle of light rays described by
$\theta,\sigma,\omega$ : the expansion, shear and vorticity. 
In the spherically symmetric case one has  $\omega=\sigma=0$ \cite{Visser:1995cc}, 
hence 
\be        \label{4}
\frac{d\theta}{d\lambda}=-R_{\mu\nu}v^\mu v^\nu-\frac12\,\theta^2\,.
\ee
If rays pass through a wormhole throat, there is a moment of minimal cross-section area, $\theta=0$
but $d\theta/d\lambda>0$, hence $R_{\mu\nu}v^\mu v^\nu<0$ and the NEC is violated. 
\fi

If the spacetime is not spherically symmetric, then the above arguments do not apply,
but there are more subtle geometric considerations showing that the wormhole throat --
a compact two-surface of minimal area -- can exist only if the NEC is violated
\cite{Friedman:1993ty,Hochberg:1998ii}.
As a result, traversable wormholes are possible only if the
energy density becomes negative, for example due to vacuum polarization
\cite{Morris:1988tu}, or due to exotic types of matter,
such as phantom fields with a negative kinetic energy
\cite{Bronnikov:1973fh,Ellis:1973yv}.

 Since energy is normally assumed to be positive, traversable wormholes
were for a long time considered to be something odd.
However, the situation changed after the discovery of cosmic acceleration
\cite{1538-3881-116-3-1009,0004-637X-517-2-565}, which prompted a large number of
alternative gravity models in which the energy is not necessarily positive definite.
Wormholes have been found
in many such theories,
as, for example,
in Gauss-Bonnet theory \cite{Maeda:2008nz,Kanti:2011jz,Cuyubamba:2018jdl},
in brane-world models
\cite{Bronnikov:2002rn},
in theories with
non-minimally coupled fields
\cite{Sushkov:2011jh}, in massive (bi)gravity \cite{Sushkov:2015fma},
in theories with classical fermions
\cite{Blazquez-Salcedo:2020czn,Konoplya:2021hsm,Dzhunushaliev:2025ntr}, 
with an additional electric field  \cite{Kim:2001ri}, 
with a cosmological constant and anisotropic matter \cite{Blazquez-Salcedo:2020nsa,Kim:2025zyo}, 
etc.
Traversable wormholes in  such theories have become
quite popular nowadays, and many people take them seriously, although it is not clear
whether alternative gravity models indeed apply to the description of physical reality.

At the same time, one should note that traversable wormholes can actually be obtained already in vacuum General Relativity,
without introducing any modifications of gravity or exotic matter, but instead allowing for line singularities in the metric
\cite{Zipoy,Bronnikov:1997gj,Clement:2015aka,Gibbons:2017jzk,Clement:2022pjr,Clement:2026pxz}. 
The singularities effectively play the role of the negative energy source needed for the wormhole existence.

\blue{
We shall call such solutions {\it vacuum wormholes}. These are 
on-shell  geometries  with several asymptotically flat regions 
that contain  distributional  matter sources. 
They are vacuum everywhere except on sets of measure zero.
}
The best-known example is the Kerr solution,
\bea \label{Kerr-m}
d{\bf s}^2=-d{\T}^2 +\frac{\Sigma}{\Delta}\,d{\R}^2+\Sigma\, d\vartheta^2 +({\R}^2+{\A}^2)\sin^2\vartheta\, d\varphi^2
+\frac{2\M\R}{\Sigma}\left( d{\T}- \A\sin^2\vartheta d\varphi\right)^2.
\eea
Here $\Sigma={\R}^2+{\A}^2\cos^2\vartheta $ and $\Delta={{\R}^2-2\M\R+{\A}^2}$, where
${\M}$ is the ADM mass and ${\A=\J/\M}$ is the ratio of the angular momentum to the mass.
Of course, this describes a black hole geometry; however,
as shown by Carter \cite{Carter:1968rr}, it also exhibits wormhole features, because it has two asymptotic regions
corresponding to the limits ${\R}\to\infty$ and ${\R}\to -\infty$.
Geodesics can interpolate between these two regions unless
they hit the curvature singularity located on the ring in the equatorial plane at
${\R}=0$, $\theta=\pi/2$, where $\Sigma=0$. This ring singularity can be viewed as a distributional
matter source carrying negative energy.

\blue{For example, 
a particle with unit mass moving along the symmetry axis avoids  the singularity and passes through the ring 
into the $\R<0$ region. Its trajectory $\R(\tau )$ is governed  by 
\be               \label{radg}
\left(\frac{d\R}{d\tau}\right)^2-\frac{2 \M \R}{\R^2+\A^2}={\cal E}^2-1. 
\ee
For any particle with energy ${\cal E}>1$, the trajectory $\R(\tau )$ extends over the entire  interval $(-\infty,\infty)$, 
thereby connecting the two asymptotic regions.
}

Let us take the ${\M}\to 0$ limit while keeping ${\A=\J/\M}$ fixed. The
Kerr line element  then reduces to \cite{Gibbons:2017djb}
\be \label{BE1}
d{\bf s}^2=-d{\T}^2+\frac{{\R}^2+\A^2\cos^2\vartheta}{{\R^2}+\A^2}\left[d{\R}^2
+({\R}^2+\A^2) d\vartheta^2\right] +({\R^2}+\A^2)\sin^2\vartheta d\varphi^2,
\ee
which is one of the Zipoy-Voorhees metrics \cite{Zipoy,Voorhees:1971wh}.
The geometry is locally flat, but it still describes a wormhole since the radial coordinate
${\R}\in (-\infty,\infty)$.
The region ${\R}>0$ covers the entire Minkowski space, as does the region ${\R}<0$,
and these two spaces are glued to each other through the wormhole throat at
${\R}=0$, which is the minimal surface
with the geometry of a disk. The
geodesics (straight lines) can cross it,
passing to the other universe \cite{Gibbons:2016bok,Gibbons:2017jzk}. 
\blue{Taking the ${\M}\to 0$ limit  in \eqref{radg}, the equation still describes unbounded motion 
interpolating  between the two asymptotic regions, with $\R(\tau)\in (-\infty,\infty)$. }

The wormhole structure persists in the ${\M}\to 0$ limit because
the ring singularity of the Kerr metric does not disappear completely, and there remains
a conical singularity of the Ricci tensor at ${\R}=\cos\vartheta=0$. This singularity can be interpreted
as a ring made of an infinitely thin cosmic string of negative tension. Therefore, the metric \eqref{BE1}
is called a ring wormhole.

\begin{figure}[th]
\hbox to \linewidth{ \hss
		\includegraphics[width=10 cm]{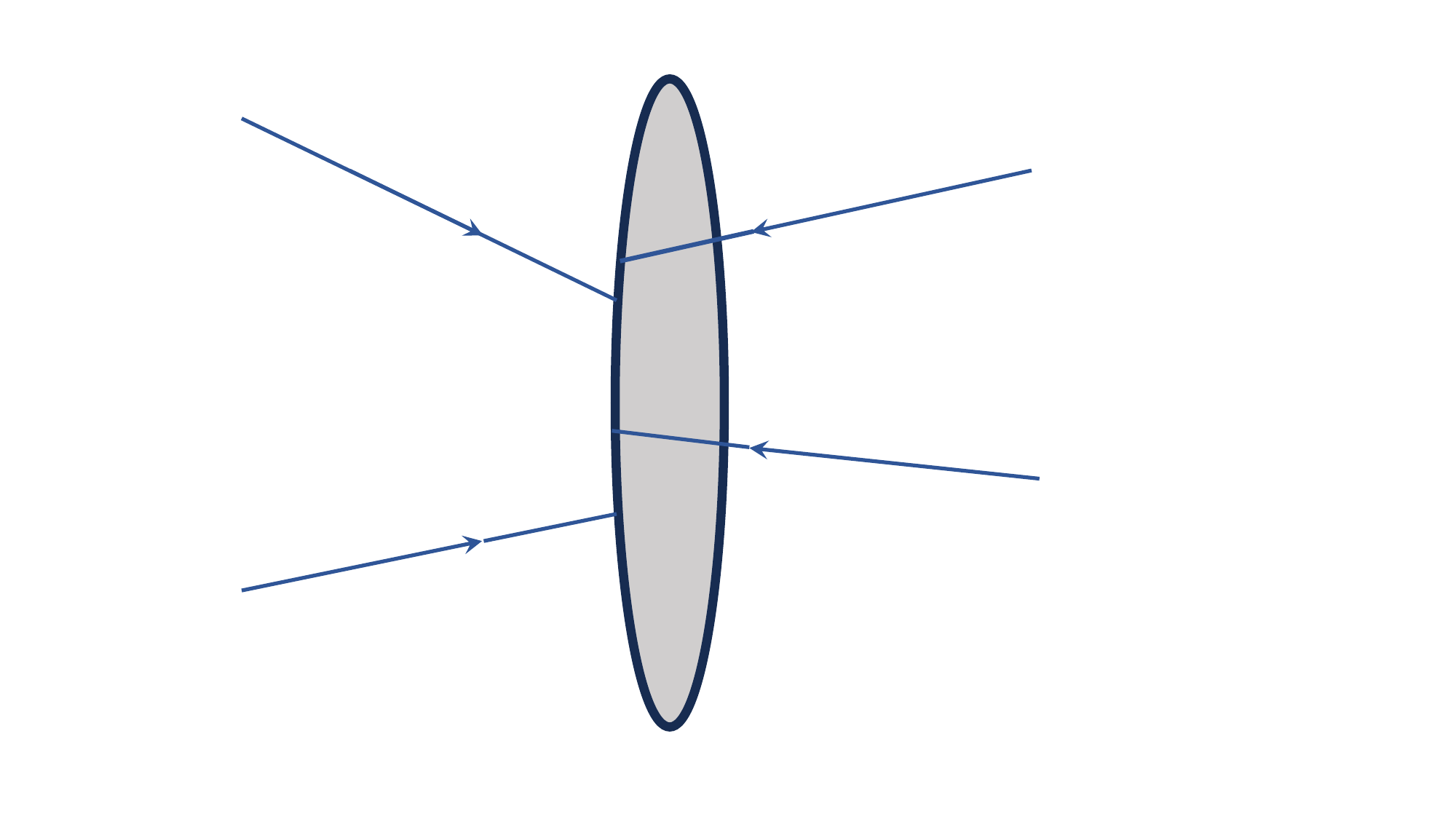}

\hss}
\caption{Particles entering the ring are not seen coming from the other side}
\label{Fig000}
\end{figure}

The static geometry \eqref{BE1} certainly represents the simplest of all possible wormholes. It can be visualized as a
``magic ring'' in space. Since the geometry is locally flat, free-falling bodies follow straight lines, and those missing
the ring stay in space, whereas those crossing the ring enter the other universe and disappear from
view (see Fig.\ref{Fig000}) \cite{Gibbons:2016bok,Gibbons:2017jzk}. However,
for this mechanism to work, the negative energy carried by the ring must be quite large \cite{Gibbons:2016bok}.
One may argue that such rings perhaps remain after the complete evaporation of Kerr black holes.

Despite its simplicity, the ring solution \eqref{BE1} has received very little attention in the literature.
For example, its stability remains an open issue. Since its geometry is locally flat, the usual laws of physics
do not change locally, but there can be global effects due to the nontrivial spacetime topology. 
A nice analysis was carried out   in Ref.\cite{Frolov:2023res}, where 
Newtonian gravity in the ring wormhole spacetime was analyzed. 
Exact solutions describing static deformations of the ring wormhole were reported in  
\cite{Sarmiento-Alvarado:2026efk}.

In what follows, we shall study stationary generalizations of this static wormhole. Of course, one such generalization
is provided by the original Kerr metric. However, the Kerr geometry is not symmetric under the
replacement ${\R}\to -{\R}$,
which swaps the two asymptotic regions, because the mass ${\M}$ then changes sign. We are instead looking for
stationary solutions that would be symmetric under ${\R}\to -{\R}$ and would have the same ADM mass when measured
from both infinities. Such solutions can be
constructed by solving the vacuum Ernst equations with suitable boundary conditions,
but they have never been described in the literature,
although various methods for solving the Ernst equations
have been widely discussed (see, e.g.
\cite{Stephani:2003tm,Belinski:2001ph}).

Our additional motivation
is provided by the fact that, by applying a simple ``dressing'' or ``scalarization''
procedure originally suggested by Eris and Gurses in a different context \cite{Eris},
the ring metric \eqref{BE1} can be transformed into the well-known
wormhole solution with a phantom scalar field found independently by Bronnikov  and Ellis (BE) \cite{Bronnikov:1973fh,Ellis:1973yv}. 
The scalarization does not change the $({\T},\varphi)$ part of the line element but removes the conformal factor
from the $({\R},\vartheta)$ part of the metric, yielding a globally regular geometry:
\be \label{BE0}
d{\bf s}^2=-d{\T}^2+d{\R}^2+({\R}^2+\A^2)(d\vartheta^2+\sin^2\vartheta d\varphi^2).
\ee
This metric is no longer vacuum but is sourced by the scalar field $\Phi=\arctan(\R/\A)$.
Therefore, the dressing removes the singularity and renders the
solution globally regular.

 The BE solution \eqref{BE0} provides a canonical example of a traversable wormhole that has been extensively studied.
In particular, there have been numerous attempts at
constructing its stationary generalizations. There are
perturbative \cite{Kashargin:2007mm,Kashargin:2008pk} and numerical
\cite{Kleihaus:2014dla,Chew:2016epf} indications in favour of the existence of a
{globally regular} spinning generalization, but
its analytical form remains unknown even now, more than 50 years after the BE discovery
\cite{Bronnikov:1973fh,Ellis:1973yv}.

 At the same time, if we stay within the vacuum theory and construct the
stationary generalization of the ring solution \eqref{BE1},
then it can be scalarized to become the
spinning BE wormhole. The advantage of this approach is that,
within the vacuum theory, one can apply various
generating methods for constructing stationary solutions. This idea was used in Ref.~\cite{Volkov:2021blw},
where a number of generating techniques were considered, but
all exact solutions obtained there are unacceptable for various reasons. At the same time,
the perturbative analysis carried out in \cite{Volkov:2021blw} suggests that a physically acceptable
stationary solution exists.

Therefore, in what follows we temporarily suspend attempts to find the stationary solution
exactly and shall instead analyze it using numerical methods. Specifically, we solve
the vacuum Ernst equations numerically, which determines the $({\T},\varphi)$ part of the line element
encoding the essential features, such as the wormhole mass and angular momentum. 
The geometry is free from NUT-type singularities along an infinite line, but its 
$({\rm r},\vartheta)$ part contains a singularity at the ring, which  can be removed
upon scalarization.

\begin{figure}[th]
\hbox to \linewidth{ \hss
	\includegraphics[width=15 cm]{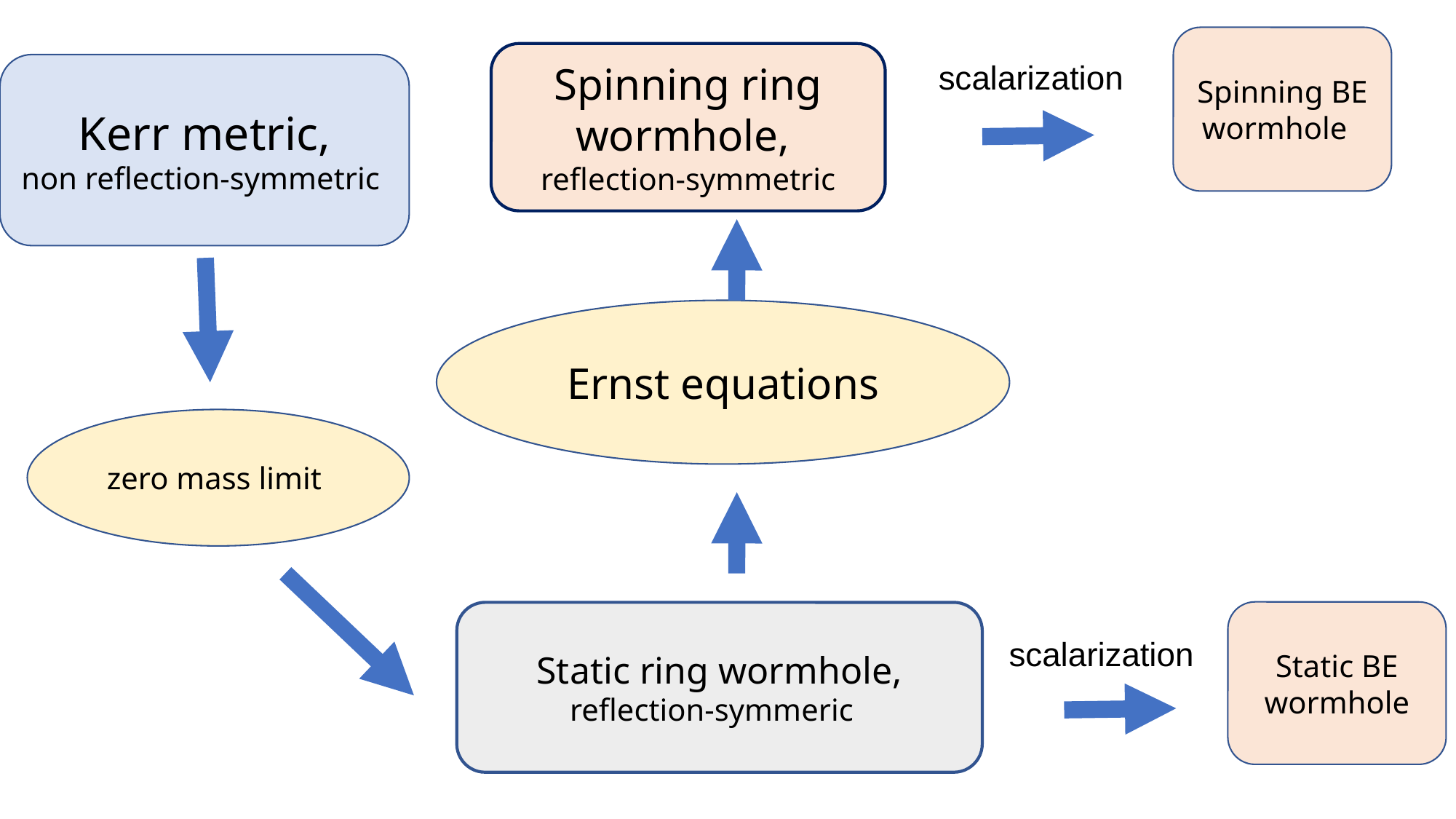}

\hss}
\caption{The Kerr metric describes a wormhole that is not symmetric under swapping the
asymptotic regions, but this symmetry is restored in the 
$M\to0$ limit, which corresponds to the 
static ring wormhole. This solution can then be spun up again, but in such a way that the reflection 
symmetry is preserved, which requires solving the vacuum Ernst equations. 
Scalarizing these vacuum solutions yields wormholes supported by a phantom scalar field.}
\label{Fig0}
\end{figure}

Our procedure
is schematically summarized by the diagram in Fig.\ref{Fig0}.
The central object is the ring wormhole obtained from the Kerr metric in the static limit.
The problem is to spin it up again, but in a way different from that of Kerr,
while respecting the symmetry under reflections across the wormhole throat.
This can be done by solving the vacuum Ernst
equations with the appropriate boundary conditions, which yields spinning ring wormholes.
If we scalarize them, this yields spinning generalizations of the BE wormhole
previously studied in \cite{Kleihaus:2014dla,Chew:2016epf} (we shall comment on our agreement
with that analysis). These solutions are globally regular,
but perhaps less
interesting physically, since already in the static limit they are
unstable \cite{Shinkai:2002gv} (although rotation may help \cite{Azad:2023iju,Azad:2024axu}).
On the other hand, spinning vacuum wormholes may be more interesting,
since they do not require exotic matter fields, are related to the Kerr metric,
and their stability properties are presently unknown.

\blue{
The stability of these vacuum wormholes is an important issue, since their physical relevance depends 
on their stability. However, a detailed stability analysis is beyond the scope of the present paper, 
and we leave this question for future work.
}

In what follows, we shall consider the vacuum theory, only briefly discussing the scalarization in Section \ref{SctIV}.
Sections \ref{SctII}, \ref{SctIII}, and \ref{SctV} describe stationary vacuum metrics, the Ernst equations, boundary conditions, and observable
parameters of the wormholes. The numerical solutions describing the wormholes and their various properties
are presented in Section \ref{SctVI}, and their scaling extensions are described in Section \ref{SctVII}.
Section \ref{SctVIII} contains a discussion of the fast-rotation limit, the geometrical properties of the wormholes
are described in Section \ref{SctIX}, and Section \ref{SctX} summarizes the results.

The appendices contain many technical details,
such as the dual formulations of the line element,
the essential facts about the Kerr metric, and the definition of the multipole moments. 

\section{The stationary vacuum gravity \label{SctII}}
\setcounter{equation}{0}

It is convenient, from the very beginning, to pass to dimensionless variables by rescaling the line element as
\be \label{mu}
d{\bf s}^2=\MU^2 g_{\mu\nu } dx^\mu dx^\nu \equiv \MU^2 ds^2,
\ee
where $\MU$ is the length scale, while $g_{\mu\nu }$ and $x^\mu$ are dimensionless. This allows us to consider only
dimensionless quantities, which will be denoted below by ordinary Greek or
Latin symbols, such as $M,J,a,\chi$. If needed,
their dimensionful versions will be
denoted by Roman symbols, such as $\M,\J,\A$.

 The vacuum gravity is determined by   the action 
\be              \label{1}
S=\frac12\, \MU^2 {\rm M}_{\rm Pl}^2\int {R}\, \sqrt{-{g}} \,d^4{x}\,, 
\ee
variation of which yields the vacuum Einstein equations,  
\be            \label{eq}
R_{\mu\nu }=0. 
\ee
We assume the spacetime   to be stationary and axially symmetric, 
\bea                \label{pf}
ds^2
=-e^{2\V}dt^2+e^{-2\V}\left( e^{2 {\gamma}}\,(d\rho^2+dz^2) +\rho^2 (d\varphi-\W dt)^2 \right), 
\eea
where $\V,\W,\gamma$ depend only on  $\rho,z$. 
This geometry  admits  two Killing vectors, $\partial_t$ and $\partial_\varphi$,  whose mutual scalar products are 
\be           \label{g00}
\left(\partial_t,\partial_t\right)&=&g_{00}=-e^{2\V}+\rho^2\W^2 e^{-2\V},~\nn \\
\left(\partial_\varphi,\partial_\varphi\right)&=&g_{\varphi\varphi}=\rho^2 e^{-2\V},~\nn \\
\left(\partial_t,\partial_\varphi\right)&=&g_{0\varphi}=-\rho^2\W e^{-2\V}. 
\ee
The Einstein equations \eqref{eq}  reduce to 
two coupled equations for $\V$ and $\W$,
\be                        \label{EEEx}
\nabla_k\nabla^k \V=\frac{\rho^2}{2} e^{-4\V} \nabla_k\W\nabla^k\W,~~~~~~~~
\nabla_k\left( \rho^2 e^{-4\V}\nabla^k \W  \right)=0, 
\ee
whose solution determines the source terms in the   first-order equations for $\gamma$, 
\be             \label{k-gamma}
\frac{1}{\rho}\,\partial_\rho {\gamma}&=&(\partial_\rho \V)^2-(\partial_z \V)^2
+\frac{\rho^2}{4}\,e^{-4\V}[(\partial_z \W)^2-(\partial_\rho \W)^2], \nn \\
\frac{1}{2\rho}\,\partial_z {\gamma}&=&\partial_\rho \V \partial_z \V 
-\frac{\rho^2}{4}\,e^{-4\V}\partial_\rho \W\, \partial_z \W\,.
\ee
Here $\nabla_k$ is the covariant derivative with respect to the flat 3-dimensional metric, 
\be
dl^2=h_{ik}dx^i dx^k=d\rho^2+dz^2+\rho^2 d\varphi^2\,,
\ee
which is also used to raise and lower the indices. 

Instead of $\W$, which we call the rotation field, it is often convenient to use the twist potential $\chioo$ defined as follows.
Using the axial Killing vector and the dual 1-form associated with it,
\be
k_{(\varphi)}^\mu \frac{\partial}{\partial x^\mu}=
\frac{\partial}{\partial\varphi},~~~~~~~k_{(\varphi)}=k_{(\varphi)\mu} dx^\mu=\rho^2 e^{-2V} (d\varphi -\W dt),
\ee
one sets
\be
k_{(\varphi)}\wedge d k_{(\varphi)}=\star (d\chioo),
\ee
where the star denotes the Hodge dual. This yields the relations
\be \label{twist}
\partial_\rho\chioo=\rho^3 e^{-4\V}\partial_z \W,~~~~~
\partial_z\chioo=-\rho^3 e^{-4\V}\partial_\rho \W,
\ee
whose integrability condition coincides with the $\W$-equation in \eqref{EEEx}.
Using this, one can eliminate $\W$ from the equations \eqref{EEEx} in favour of $\chioo$, thus obtaining
\be \label{EEExx}
\nabla_k\nabla^k \V=\frac{e^{4\V} }{2\rho^4}\, \nabla_k\chioo\nabla^k\chioo,~~~~~~~~
\nabla_k\left( \frac{e^{4\V}}{\rho^4} \nabla^k \chioo \right)=0.
\ee
These two equations can be combined into a single equation for the
complex-valued Ernst potential ${\cal E}_{(\varphi)}=\rho^2 e^{-2\V}+i\chioo$ (see Eq.\eqref{Er} in \ref{AppA})
\cite{Ernst}.
However, we shall use the equations in the real form \eqref{EEExx}, or in the form \eqref{EEEx},
while always referring to them as the Ernst equations.

\section{Wormholes \label{SctIII}}
\setcounter{equation}{0}

To describe wormholes, one passes  to the spheroidal coordinates $x,y$ defined by 
\be               \label{rz}
\rho=\sqrt{(x^2+a^2)(1-y^2)},~~~~z=xy.
\ee
Here $a$ is a parameter, 
one has $x\in(-\infty,\infty)$ and $y\equiv \cos\vartheta \in [-1,1]$. The spacetime metric \eqref{pf} assumes the form 
\be               \label{METR}
ds^2
=-e^{2\V}dt^2+e^{-2\V} \left(e^{2 {\K}}\left[dx^2+\frac{x^2+\nuu }{1-y^2}\, dy^2\right] +(x^2+\nuu )(1-y^2)
 (d\varphi-\W\, dt)^2\right),~~~~
\ee
where $\V,\W,\K$ now depend on  $x,y$, and 
\be                \label{eK}
e^{2{\K}}=\frac{x^2+\nuu\,  y^2}{x^2+\nuu }\, e^{2{\bm\gamma}}. 
\ee
The limits $x\to\pm\infty$ correspond to the asymptotic regions, where $\V,\W,\K$  approach zero. 
The asymptotic regions are connected through the wormhole throat at 
$x=0$. The equatorial section  of the throat at 
$x=y=0$  is a circle of  radius 
\be             \label{R0}
R_0=\left.a\,e^{-\V}\right|_{x=y=0}, 
\ee
which extends along the azimuthal $\varphi$-direction. As will be explained below, this circle supports
a  curvature singularity. 
\blue{One should emphasize that $a$ is a coordinate parameter, which we shall refer to as the size parameter, 
whereas $R_0$ is the geometrical circumferential radius, which we shall call the ring radius or wormhole size. 
These two parameters coincide in the static limit, when $\V=\W=0$ and hence $R_0=a$. 
Thus, $a$ represents the geometrical size of the static wormholes. 
For spinning configurations, one has $R_0>a$, since the wormholes are stretched by the centrifugal force. 
The corresponding dimensionful value of $R_0$ is obtained by 
multiplying it by the length scale, ${\rm R}_0=\MU R_0$.
}

The symmetry axis is located at $\rho=0$, corresponding to the set invariant under the action of the
axial Killing vector $\partial/\partial\varphi$. Interestingly,
wormholes actually possess two symmetry axes, because 
the condition $\rho=0$ is achieved either for $y=1$, $z=x\in(-\infty,\infty)$, or for $y=-1$, $z=-x\in(-\infty,\infty)$
(see Fig.\ref{Fig1oo} below). It is worth emphasizing that the amplitude $\K$ must vanish on the axes 
in order to avoid a conical
singularity there. One therefore requires
\be                  \label{Kaxe}
\left.\K\right|_{y=\pm 1}=0.
\ee
An additional condition to be imposed on the axes is that
$g_{0\varphi}=-\rho^2 W e^{-2\V}$
must vanish there as well in order to avoid a Taub--NUT-type
singularity \cite{Taub:1950ez}. This condition is automatically satisfied provided that $\W$ and $\V$ remain bounded everywhere.

Let us now consider the field equations expressed in spheroidal coordinates.

\subsection{Ernst equations}

Introducing the notation 
\be
\langle f,g\rangle\equiv (x^2+\nuu )f_{,x}g_{, x}+(1-y^2)f_{, y} g_{,y}\,,
\ee
where the comma denotes the partial derivative, 
equations \eqref{EEEx} assume the form 
\be                     
\label{UUU}
&&[(x^2+\nuu )\V_{,x}]_{,x}+[(1-y^2)\V_{,y}]_{,y}-\frac12\,\rho^2 e^{-4\V}\langle \W,\W\rangle=0,~~~ \nn \\
&&\left[\rho^2\,e^{-4\V}\,(x^2+\nuu )\,\W_{,x}\right]_{,x}+\left[\rho^2\,e^{-4\V}\, (1-y^2)\, \W_{,y}\right]_{,y}=0.
\ee
These  equations 
can be obtained by varying the functional 
\be                    \label{Ew}
E_{\W}=\int\left( \langle \V,\V\rangle-\frac{\rho^2}{4}\,e^{-4\V}\langle \W,\W\rangle \right) dx \,dy\,.
\ee
Solutions for $\V,\W$ determine the $g_{00},g_{0\varphi},g_{\varphi\varphi}$ metric components in \eqref{g00},
thus determining  the ADM mass, angular momentum, and the higher multipole moments.

It will be also useful to consider the equations in the $\V,\chioo$ form \eqref{EEExx}. 
The relations \eqref{twist} between the rotation field $\W$ and  the twist potential  $\omega$ now read 
\be              \label{twisto}
\W_{, x}=\frac{y^2-1}{\rho^4}\, e^{4\V} \chioo_{, y}\,,~~~~~~~~
\W_{, y}=\frac{x^2+\nuu }{\rho^4}\, e^{4\V} \chioo_{, x}\,,
\ee
and the Ernst equations  \eqref{EEExx} become 
\be                     
\label{UUU1}
&&[(x^2+\nuu )\V_{,x}]_{,x}+[(1-y^2)\V_{,y}]_{,y}-\frac{1}{2\rho^4} e^{4\V}\langle \chioo,\chioo\rangle=0,~~~ \nn \\
&&\left(\frac{x^2+\nuu }{\rho^4}\,e^{4\V}\,\chioo_{,x}\right)_{,x}+\left(\frac{1-y^2}{\rho^4}\,e^{4\V}\, \chioo_{,y}\right)_{, y}=0.
\ee
These  can be obtained by varying the functional 
\be                 \label{Ec}
E_{\chioo}=\int\left( \langle \V,\V\rangle+\frac{1}{4\rho^4}\,e^{4\V}\langle \chioo,\chioo\rangle \right) dx \,dy\,.
\ee

In what follows, we shall study  the Ernst  equations both in the $\V,\W$ form \eqref{UUU} and in the 
 $\V,\chioo$ form \eqref{UUU1}. 
 Both forms yield equivalent results, but the numerical 
 procedure for the 
  $\V,\chioo$-equations turns out to be more efficient and converges faster, 
  presumably because the underlying functional $E_{\chioo}$ is positive 
  definite.

\subsection{Equations for the $\K$-implitude}

Solving the Ernst equations for $\V$ and $\W$ determines the $g_{00},g_{0\varphi},g_{\varphi\varphi}$ metric 
components.  The remaining metric components $g_{xx}$ and $g_{yy}$ depend  also on the amplitude $\K$. 
Using \eqref{k-gamma} and \eqref{eK}, one obtains the equations 
\be               \label{KK0}
\partial_x \K&-&\frac{1-y^2}{x^2+\nuu \,y^2}\left( \Gamma(\V)
-\frac{1}{4}\,(x^2+\nuu )(1-y^2) \,e^{-4\V}\,\Gamma(\W)+\frac{\nuu \, x}{x^2+\nuu }
\right)=0, \nn \\
\partial_y \K&-&\frac{x^2+\nuu }{x^2+\nuu \,y^2}\left( \Lambda(\V)
-\frac{1}{4}\,(x^2+\nuu )(1-y^2)\, e^{-4\V}\,     \Lambda(\W)+\frac{\nuu \, y}{x^2+\nuu }
\right)=0, 
\ee
where  the following abbreviations are introduced, 
\be               \label{SL}
\Gamma(f)&\equiv& x(x^2+\nuu ) f_{,x}^2-2y(x^2+\nuu )f_{,x} f_{,y}+x(y^2-1) f_{,y}^2, \nn \\
\Lambda(f)&\equiv& y(x^2+\nuu ) f_{,x}^2+2x(1-y^2)f_{,x} f_{,y}+y(y^2-1) f_{,y}^2\,. 
\ee
The consistency condition for these equations, $\partial_x\partial_y\K=\partial_y\partial_x\K$, is fulfilled by virtue
of the Ernst equations for $\V,\W$. 
The amplitude  $\K$  can be split as 
\be
\K=\K_{\rm reg}+\K_{\rm sing},
\ee
where $\K_{\rm reg}$ is bounded, while $\K_{\rm sing}$ diverges for $x\to 0$, $y\to 0$. 
This singularity  is partly sourced by the  terms in \eqref{KK0} which do not depend on $\V,\W$: 
\be                     \label{3.14}
\partial_x \K_{\rm sing}&=&\frac{1-y^2}{x^2+\nuu \,y^2}\times \frac{\nuu \, x}{x^2+\nuu }+\ldots,~~~~~
\partial_y \K_{\rm sing }=\frac{x^2+\nuu }{x^2+\nuu \,y^2}\times \frac{\nuu \, y}{x^2+\nuu }+\ldots ,
\ee
integrating which yields 
\be
\K_{\rm sing}=\frac12\ln\frac{x^2+\nuu\,  y^2}{x^2+\nuu }+\ldots. 
\ee
In addition, 
some of the  $\V,\W$-dependent terms in \eqref{KK0} also give a contribution 
which  diverges for $x\to 0$, $y\to 0$. The $\V$-contribution is finite, 
since, as will be explained  below, one has  $\V_x=\V_y=0$  at $x=y=0$. 
However, the $\W$-contribution is singular, because 
$\W_x$  does not  vanish  at $x=y=0$. 

Combining  together the singular terms from \eqref{3.14} and those determined by  $\W_x$, 
the total contribution to the singular  part of the $\K$-amplitude  is determined by 
\be                \label{Ksing}
\partial_x \K_{\rm sing}=\eta\times \frac{1-y^2}{x^2+\nuu \,y^2}\times \frac{\nuu \, x}{x^2+\nuu },~~~~~~~~~
\partial_y \K_{\rm sing }=\eta\times \frac{x^2+\nuu }{x^2+\nuu \,y^2}\times \frac{\nuu \, y}{x^2+\nuu }, 
\ee
integrating which yields 
\be              \label{Kr2oo}
\K_{\rm sing}=\frac\eta2\ln\frac{x^2+\nuu\,  y^2}{x^2+\nuu }. 
\ee
Here we have introduced the ``tension parameter'' and used \eqref{R0}:
\be                \label{Kr2o}
\C\equiv 1-\left.\frac{a^4}{4}\, e^{-4\V}\W_{,x}^2\right|_{x=y=0}=1-\left.\frac{R_0^4}{4}\,\W_{,x}^2\right|_{x=y=0}.
\ee
Subtracting the singular contribution, the remaining part of the $\K$ amplitude is regular and determined by 
\be               \label{KKr}
\partial_x \K_{\rm reg}&=&\frac{1-y^2}{x^2+\nuu \,y^2}\left( \Gamma(\V)
-\frac{1}{4}\,(x^2+\nuu )(1-y^2) \,e^{-4\V}\,\Gamma(\W)+(1-\eta)\times \frac{\nuu \, x}{x^2+\nuu }
\right), \nn \\
\partial_y \K_{\rm reg}&=&\frac{x^2+\nuu }{x^2+\nuu \,y^2}\left( \Lambda(\V)
-\frac{1}{4}\,(x^2+\nuu )(1-y^2)\, e^{-4\V}\,     \Lambda(\W)+(1-\eta)\times \frac{\nuu \, y}{x^2+\nuu }
\right). ~~~~
\ee
Requiring that
$\K_{\rm reg}\to 0$ as $x\to\infty$ implies that
$\K_{\rm reg}$ vanishes everywhere on the symmetry axes,
since $\partial_x \K_{\rm reg}=0$ for $y=\pm 1$. The singular part $\K_{\rm sing}$ in \eqref{Kr2oo}
also vanishes there, and therefore the regularity condition \eqref{Kaxe} for the amplitude $\K$ on the axes is automatically satisfied.

\subsection{Wormhole source}
Suppose that we find a globally   regular solution of the Ernst equations for which  $\V,\W$ approach zero for $x\to\pm\infty$. 
Computing the tension parameter  $\eta$  in \eqref{Kr2o}, we  can then 
determine from  \eqref{Kr2oo} and \eqref{KKr} the amplitudes $\K_{\rm sing}$ and 
$\K_{\rm reg}$. 
The resulting solution is 
\be               \label{METR1}
ds^2
&=&-e^{2\V}dt^2+{\cal K}\times e^{-2\V+2\K_{\rm reg}}
\left[dx^2+\frac{x^2+\nuu }{1-y^2}\, dy^2\right] +e^{-2\V}\rho^2\, (d\varphi-\W\, dt)^2\,,~~~~\nn \\
{\cal K}&\equiv&e^{2\K_{\rm sing}}=\left(\frac{x^2+\nuu\,  y^2}{x^2+\nuu }\right)^\C. 
\ee
The ${\cal K}$-factor vanishes at $x=y=0$, producing  a curvature singularity 
along the line in the equatorial plane that encircles 
the wormhole throat in the azimuthal direction. This singularity can be interpreted as 
a distributional matter source that   carries the negative energy 
needed for the wormhole existence. 
Specifically, denoting 
\be
C=\left.e^{\K_{\rm reg}-\V}\right|_{x=y=0}, 
\ee 
the $(x,y)$ part of the line element  \eqref{METR1} reduces in the vicinity of $x=y=0$ to 
\be                 \label{cone}
C^2\left(x^2+\nuu  y^2\right)^\C(dx^2+\nuu  dy^2)=C^2 u^{2\C}(du^2+u^2d\phi^2)=d{\varrho}^2+{\varrho}^2\, d\psi^2\,,
\ee
where the following coordinate transformations have been used, 
\be
x=u\cos\phi, ~~~a\,y=u\sin\phi, ~~~~~\varrho=C \frac{u^{\C+1}}{(\C+1)}, ~~~~~~\psi=(\C+1)\phi
\ee
($\phi$ is not the azimuthal angle $\varphi$ in \eqref{METR1}). 
The line element in \eqref{cone} is the flat 2-metric expressed in polar coordinates
$\varrho,\psi$. 
However, since $\phi\in[0,2\pi]$, one has 
$\psi\in[0,2\pi +2\C\pi]$.  Therefore,  \eqref{cone} describes the geometry of 
a cone with a negative angle deficit of $\Delta \psi= -2\C\pi$. 
This conical singularity can be interpreted as a a 
distributional matter source, which can be described as  a cosmic string 
of {\it negative} tension, 
\be           \label{TT}
T=-\frac{\C}{4}, 
\ee
extending along the azimuthal $\varphi$-direction  
\cite{Gibbons:2016bok,Gibbons:2017jzk}. In other words, 
this is a ring made of an infinitely thin  cosmic string encircling the wormhole throat. 
Therefore, the vacuum wormhole is created by a ring of negative tension $T$  and of radius $R_0$ 
determined by \eqref{R0}.

 Restoring the correct physical dimensions for a moment, the tension of a cosmic string, $\mathrm{T}$, 
 that is, the energy per unit length,
is related to the angle deficit via $\Delta\psi=-2\pi\eta=(8\pi \mathrm{G/c^4})\mathrm{T}$. 
This gives the dimensionful analogue
of \eqref{TT},
\be               \label{TTT}
\mathrm{T}=-\frac{\C}{4}\times\frac{c^4}{G}.
\ee

\blue{
The structure of the distributional matter source can also  be determined  from   the field equations. 
Since the amplitudes $\V,\W$ are smooth and bounded, 
the source is associated with derivatives of the amplitude $K$, which contains  a singular  part 
$\K_{\rm sing}$. 
\\
\indent
Let us decompose  the metric as 
$g_{\mu\nu}=\eta_{\hat{a}\hat{b}}\,\Theta^A_\mu\Theta^B_\nu$ where $\eta_{AB}=\diag [-1,1,1,1]$ and 
\be                   \label{Theta}
\Theta^0=e^{\V} dt,~~\Theta^1=e^{\K-\V} dx,~~\Theta^2=e^{\K-\V}\sqrt{ \frac{x^2+a^2}{1-y^2}}\, dy,~~
\Theta^3=\rho\,e^{-V}(d\varphi-W\, dt),~~~~~
\ee
with $\rho=\sqrt{x^2+a^2)(1-y^2)}$. Introducing the dual tetrad   $e_B^{~\mu}$  such that 
$\Theta^A_{~\mu}\,e_B^{~\mu} =\delta^A_{~B}$,   the non-zero tetrad projections 
of the Ricci tensor are $R_{\hat{0}\hat{0}}=R_{\hat{3}\hat{3}}$, $R_{\hat{0}\hat{3}}$, $R_{\hat{1}\hat{1}}$, 
$R_{\hat{2}\hat{2}}$, 
$R_{\hat{1}\hat{2}}$. 
These determine the four equations in \eqref{UUU} and \eqref{KK0} considered above, 
\be             \label{EEEK1}
EqV&=& e^{2\K-2\V}\,(x^2+a^2)\,R_{\hat{0}\hat{0}} =0,~\nn \\
EqW&= &2\,e^{2\K-4\V}\,\rho\,(x^2+a^2)\,R_{\hat{0}\hat{3}}=0,~~~\nn \\
EqK_x&= &-\frac{\rho^2\, e^{2\K-2\V}}{2\,(x^2+a^2 y^2)}\,
\left( x\,(R_{\hat{1}\hat{1}}-R_{\hat{2}\hat{2}})+2\,y\,\sqrt{\frac{x^2+a^2}{1-y^2}}\,R_{\hat{1}\hat{2}}\right)=0,   \nn \\
EqK_y& = &-\frac{\rho\,(x^2+a^2) e^{2\K-2\V}}{2\,(x^2+a^2 y^2)}\,
\left( 2\,x\,R_{\hat{1}\hat{2}} -\,y\,\sqrt{\frac{x^2+a^2}{1-y^2}}\,(R_{\hat{1}\hat{1}}-R_{\hat{2}\hat{2}})\right)=0.
\ee
Here the symbols on the left-hand side denote the expressions appearing  on the left-hand-sides of \eqref{UUU}
and in \eqref{KK0}. For example, $EqV=[(x^2+\nuu )\V_{,x}]_{,x}+\ldots$, 
$EqK_x=\partial_x K-\ldots$, etc. 
\\
\indent
The singular source does not appear explicitly in these
equations because  they contain only first derivatives of $K$. 
However, there is one more  equation, 
\be                \label{EEEK2}
R_{\hat{0}\hat{0}}-R_{\hat{1}\hat{1}}&=& e^{2\V-2\K}\left( 
\K_{,xx}+\frac{[(1-y^2)\K_{,y}]_{,y}}{x^2+a^2}
+2\,\V_{,x}^2
+\frac{2\,a^2}{(x^2+a^2)^2}
\right) \nn \\
&-&e^{-2\K-2\V}(1-y^2)\left(
(x^2+a^2)\,\W_{,x}^2
+
\frac12\,(1-y^2)\, \W_{,y}^2
\right)=0. 
\ee
This relation  is a differential consequence of the equations in \eqref{EEEK1}. 
However,  it contains second derivatives of $\K$, whose singular part gives rise to  a Dirac delta function.
Consequently, the right-hand side of \eqref{EEEK2} is not zero in the distributional sense, but contains a delta function.
\\
\indent
The amplitude  $\K$ diverges at $x=y=0$, and close to this point one has 
\be              \label{Kr2ooo}
\K=\frac\eta2\ln\left( x^2+\nuu\,  y^2\right)+\ldots ,
\ee
whereas 
\be
R_{\hat{0}\hat{0}}-R_{\hat{1}\hat{1}}&=&e^{2\V-2\K}\left(\K_{,xx}+\frac{1}{a^2}\,\K_{,yy}  \right) +\ldots,
\ee
where the dots denote subleading terms. Using the well-known formula, 
\be
\left(\partial^2_{xx}+\partial^2_{yy} \right) \frac12\ln\left(x^2+y^2\right)=2\pi\,\delta(x)\delta(y), 
\ee
one concludes that 
\be
R_{\hat{0}\hat{0}}-R_{\hat{1}\hat{1}}=e^{2\V-2\K}\times 2\pi \eta\,\times  \delta(x)\delta(ay). 
\ee
Since $R_{\hat{0}\hat{0}}=R_{\hat{3}\hat{3}}=0$ by virtue of  \eqref{EEEK1}, and since 
$R_{\hat{1}\hat{1}}-R_{\hat{2}\hat{2}}=R_{\hat{1}\hat{2}}=R_{\hat{0}\hat{3}}=0$, 
it follows that the distributional contribution  is contained  in 
\be           \label{RG}
R_{\hat{1}\hat{1}}=R_{\hat{2}\hat{2}}=G_{\hat{0}\hat{0}}=-G_{\hat{3}\hat{3}}=e^{2\V-2\K}\times 2\pi \eta\,\times  \delta(x)\delta(ay). 
\ee
Using the Einstein equations $G_{AB}=8\pi T_{AB}$, one obtains the non-zero components of the  energy-moment tensor, 
\be                    \label{T25}
T_{\hat{0}\hat{0}}=-T_{\hat{3}\hat{3}}=-e^{2\V-2\K}\times \frac{\eta}{4}\times  \delta(x)\delta(ay). 
\ee
Integrating over $x,y$ yields the energy per unit length in the orthogonal (azimuthal) direction, 
\be
T=\int T_{\hat{0}\hat{0}}\,\sqrt{g_{xx}\, g_{yy}}\, dx\, dy
=\int T_{\hat{0}\hat{0}}\, e^{2\K-2\V} \sqrt{ \frac{x^2+a^2}{1-y^2}} \, dx\, dy=-\frac{\eta}{4}.
\ee
This reproduces the formula \eqref{TT} for the string tension. 
\\
\indent
The  quantities $T_{\hat{0}\hat{0}}$ and $T_{\hat{3}\hat{3}}$ in \eqref{T25} are tetrad projections. 
Passing to the space-time components, the distributional matter source supporting the wormhole is 
\be               \label{T26}
T_{\mu\nu}=e^{2\V-2\K}\, \frac{\eta}{4}\,\left(-\Theta^0_\mu\Theta^0_\nu + \Theta^3_\mu\Theta^3_\nu\right) \,  \delta(x)\delta(ay). 
\ee
This  describes  a ring formed by a cosmic string with  negative energy and tension $T=-\eta/4$. 
The only non-zero  space-time components are $T_{00}$, $T_{\varphi\varphi}$ and $T_{0\varphi}$, with  the latter 
vanishing  in the static limit. 
}
 
\subsection{Static limit}

The static limit corresponds to the trivial solution of the Ernst equations, $\V=\W=0$, 
in which case $\K_{\rm reg}=0$ and  $\C=1$, such that   the line element  \eqref{METR1} reduces to 
\be                 \label{ring}
ds^2=-dt^2+ \frac{x^2+\nuu\,y^2}{x^2+\nuu}\,\left[dx^2+\frac{x^2+\nuu}{1-y^2}\,dy^2\right]+(x^2+\nuu)(1-y^2) d\varphi^2.
\ee
Setting   $t={\rm t}/\MU$, $x={\rm r}/\MU$, $a={\rm a}/\MU$, $y=\cos\vartheta$ and   
multiplying by the scaling factor  $\MU^2$ yields the ring wormhole metric \eqref{BE1}. 

Passing to the cylindrical coordinates 
with the same transformation as in \eqref{rz}, but performed separately on each  wormhole side, such that 
$\rho_\pm=\sqrt{(x^2+\nuu)(1-y^2)}$, $z_\pm=xy$,
where the signs correspond, respectively, to $x>0$ and $x<0$, the line element reduces to 
\be
ds^2=-dt^2+d\rho_\pm^2+dz_\pm^2+\rho^2_\pm d\varphi^2\,.
\ee
 This describes two copies of Minkowski space glued to each other through the disk at $x=z_\pm=0$, 
 where $\rho\equiv \rho_\pm\in[0,1]$ and $\varphi\in [0,2\pi)$ (see \cite{Gibbons:2017jzk} for details).
This disk is the wormhole throat. The boundary of the disk
is the unit circle $\rho=1$, $\varphi\in [0,2\pi)$. It carries the conical singularity with the angle deficit of $-2\pi$,
corresponding to a ring of tension $T=-1/4$. Therefore, the geometry \eqref{ring} is locally flat but the
topology is non-trivial
and corresponds to a wormhole connecting two flat universes through the disk
\cite{Gibbons:2016bok,Gibbons:2017jzk}.

It is worth emphasizing that the ring radius $a=R_0$ can be arbitrary;
hence Eq.\eqref{ring} actually describes not one ring but
a family of rings of size $a\in (0,\infty)$. If the dimensionful ring radius is $\mathrm{R}_0=\MU\, a=1$ meter, say,
then, using \eqref{TTT}, the ring energy is
$\mathrm{E}_{\rm ring}=2\pi \mathrm{R}_0 \mathrm{T}=-\pi\mathrm{c^4}/(2 \mathrm{G})
\approx -2\times 10^{27}\,\mathrm{kg}\times \mathrm{c}^2$,
which is close to the mass of Jupiter, up to a sign  \cite{Gibbons:2016bok}.

 \blue{
In the static limit, the curvature singularity at the ring is purely distributional and is described 
by the delta-like behavior of the Ricci tensor in \eqref{RG}. For the stationary solutions, 
the singularity contains both a distributional part localized at the ring and a volume part 
associated with components of the Riemann tensor that diverge as $x,y\to0$. 
The latter divergence arises because  the 
tension parameter $\eta$ differs from the static value $\eta=1$. 
For example, setting, for simplicity,  all regular amplitudes 
 in the line element \eqref{METR1} to zero, 
\be
\V=\W=\K_{\rm reg}=0:~~~~~R_{\mu\nu\alpha\beta}R^{\mu\nu\alpha\beta}
=\frac{12\,a^2\, (1-\eta)^2 }{(x^2+a^2)^4}\times \left(\frac{x^2+\nuu}{x^2+a^2 y^2} \right)^{2\eta} \,,
\ee
and hence the curvature diverges for $x,y\to0$. 
In the static limit one has  $\eta\to 1$ and   the curvature vanishes everywhere outside the ring, 
leaving only the delta-like singularity of the Ricci tensor located at the ring.
}

 The presence of a singularity in the vacuum solutions is necessary, because it imitates the negative energy
needed for the wormhole existence. On the other hand, as we shall now see, explicitly including a negative
energy source renders the singularity unnecessary, and the wormholes become globally regular.

\section{Scalarization \label{SctIV}} 
\setcounter{equation}{0}

It is possible to include a matter source in the theory such that the vacuum solutions remain essentially the same,
but the singularity in the metric disappears. Specifically, extending the vacuum theory by
adding a real scalar field with a ``wrong" sign in front of the kinetic term changes the action \eqref{1} to
\be              \label{1a}
S=\frac12\, \MU^2 {\rm M}_{\rm Pl}^2\int \left( {R}+2\partial_\mu\Phi\partial^\mu\Phi  \right) \sqrt{-{g}} \,d^4{x}\,. 
\ee
The field equations  become 
\be           \label{eqa}
R_{\mu\nu }+2\partial_\mu\Phi\partial_\nu \Phi=0,~~~~~~\nabla_\mu \nabla^\mu\Phi=0. 
\ee
Choosing the metric in the same form as before and assuming
the scalar field to be independent of $t,\varphi$, one still has
$R_{00}=0$, $R_{0\varphi}=0$, $R_{\varphi\varphi}=0$. As a result, the equations containing the $\V,\W$
metric amplitudes -- the Ernst equations -- remain the same as in the vacuum case.

However, including the scalar field modifies 
the $\K$-amplitude. \blue{
The scalar field satisfies 
\be                           \label{scalareq}
[(x^2+\nuu )\Phi_{,x}]_{,x}+[(1-y^2)\Phi_{,y}]_{,y}=0,
\ee
which can be solved by setting $\Phi=X_l(x)P_l(y)$, where $P_l(y)$ are the Legendre polynomials, while 
\be
\left[(x^2+a^2)X_l^\prime\right]^\prime=l(l+1) X_l\,. 
\ee
This equation can be solved  analytically (see Sec.V in \cite{Volkov:2021blw} for details). 
It turns out  that  $X_l(x)$  is everywhere bounded 
for $l=0$, whereas for $l>0$ it grows at infinity as $x^l$. Therefore, requiring the solutions 
to be bounded eliminates all modes with $l>0$. 
Thus,  the only bounded solution for the scalar field corresponds to  $l=0$, 
\be              \label{Phio}
\Phi=C_\phi\, \arctan\left(\frac{x}{a} \right).
\ee
}
Using  this field as the source to the Einstein equations in \eqref{eqa} modifies 
the right-hand-side of equations 
\eqref{KK0}  for  the $\K$-amplitude, leading  to the 
replacement 
$
\K\to \K+\K_{\Phi}, 
$
where 
\be
\partial_x \K_\Phi=-C^2_\phi\times \frac{1-y^2}{x^2+\nuu \,y^2}\times \frac{\nuu \, x}{x^2+\nuu },~~~~~~
\partial_y \K_\Phi=-C^2_\phi\times \frac{x^2+\nuu }{x^2+\nuu \,y^2}\times \frac{\nuu \, y}{x^2+\nuu }. 
\ee
These equations have exactly the same structure as equations  \eqref{Ksing} for the singular part $\K_{\rm sing}$. 
As a result, the only effect  of  the scalar field is to modify the singular part of the $\K$-amplitude, 
\be              \label{Kphi}
\K_{\rm sing}\to \K_{\rm sing}+\K_\Phi=\frac{\eta-C^2_\phi}{2}\,\ln\frac{x^2+\nuu  y^2}{x^2+\nuu }.
\ee
Remarkably, this can be set to zero by choosing the integration constant $C_\phi$ for the scalar field such that 
\be               \label{Cp}
C_\phi=\sqrt{\eta}, 
\ee 
which  removes the singular factor ${\cal K}$ from the metric. 

Therefore, including the scalar field converts the vacuum wormholes into globally regular solutions. We shall call them 
scalarized or scalar-dressed wormholes. The stationary dressed wormhole is described  by the same metric as in \eqref{METR1},
but without the singular factor, 
 \be               \label{METR2}
ds^2
&=&-e^{2\V}dt^2+e^{-2\V+2\K_{\rm reg}}
\left[dx^2+\frac{x^2+\nuu }{1-y^2}\, dy^2\right] +e^{-2\V}\rho^2\, (d\varphi-\W\, dt)^2\,,~~~~\nn \\
\Phi&=&\sqrt{\eta}\,\arctan\left(\frac{x}{a} \right). 
\ee
The amplitudes $\V,\W,K_{\rm reg}$ are the same as in \eqref{METR1}, while the $g_{00}$, $g_{0\varphi}$, $g_{\varphi\varphi}$
metric coefficients are also unchanged; hence the ADM mass and angular momentum are the same as in the vacuum theory.
It follows that the very existence of the wormhole and all its essential features are determined by the vacuum theory;
the only role of the scalar field is to remove the singular factor from the
$x,y$ part of the line element.

In the static limit, \eqref{METR2} reduces to 
\be                 \label{ring1}
ds^2=-dt^2+dx^2+\frac{x^2+\nuu}{1-y^2}\, dy^2+(x^2+\nuu)(1-y^2)\,d\varphi^2,~~~~~~
\Phi=\arctan\left(\frac{x}{a}\right).
\ee
Multiplying the metric by the scale  factor $\MU^2$ and  setting  
$t={\rm t}/\MU$, $x={\rm r}/\MU$, $a={\rm a}/\MU$, $y=\cos\vartheta$
yields the BE solution  \eqref{BE0}. 

\blue{
Of course, the global regularity of the dressed solutions is appealing, but it comes at the cost of introducing 
a phantom field whose physical origin is unclear. The regularity is thus achieved at the expense of an additional 
unphysical field. Moreover, the presence of the phantom field leads to an instability already in the static limit \cite{Shinkai:2002gv}. At the same time, the phantom field is not necessary for the existence of the wormholes -- 
wormholes already exist in the vacuum theory and share all the essential features of their dressed counterparts. 
We therefore prefer not to introduce additional unphysical fields and return to the vacuum theory from now on. 
The dressed configurations will be invoked only to illustrate features associated with the $\K$-amplitude, 
which differ from those of the vacuum solutions.
}

\section{Ernst equations -- boundary conditions and conserved quantities \label{SctV}}
\setcounter{equation}{0}

Let us consider the Ernst equations \eqref{UUU}. 
We require their 
solutions to be asymptotically flat, such  that for $x\to\pm \infty$  one has 
\be         \label{41}
-g_{00}&=&e^{2\V}-\rho^2 \W^2 e^{-2\V}=1-\frac{2M}{|x|}+\ldots,~~~\nn \\
-g_{0\varphi}&=&\rho^2 \W e^{-2\V}=\frac{2J\sin^2\vartheta}{x}+\ldots ,\nn \\
g_{\varphi\varphi}&=&\rho^2 e^{-2\V}=\left(x+{\cal O}\left(1 \right) \right)^2\sin^2\vartheta, 
\ee
where the dots denote   subleading terms. 
Therefore, one should have 
\be               \label{flat}
\V=-\frac{M}{|x|}+\ldots,~~~~\W=\frac{2J}{x^3}+\ldots ~~~~~\text{as}~~~~x\to\pm \infty. 
\ee
The Schwarzschild radial coordinate $r_{\rm S}$ is defined by the condition 
\be                     \label{Schw}
g_{\varphi\varphi}=r_{\rm S}^2\sin^2\vartheta,~~~~~r_{\rm S}>0,
\ee
hence the last relation in \eqref{41} implies that for  $x\to\pm\infty$ one has 
\be
|x|=r_{\rm S}+{\cal O}\left(1 \right),~~~~~~x=\pm r_{\rm S}+{\cal O}\left(1 \right). 
\ee
Therefore, one obtains  for  $x\to\pm\infty$ 
\be
-g_{00}=1-\frac{2M}{r_{\rm S}}+\ldots,~~~
-g_{0\varphi}=\pm \frac{2J\sin^2\vartheta}{r_{\rm S}}+\ldots . 
\ee
The first of these relations 
 shows that the parameter $M$ is the ADM mass, and it is the same when measured from both infinities. 
On the other hand, $+J$ is the angular momentum measured from the $x\to+\infty$ region, while 
$-J$ is the angular momentum measured at $x\to-\infty$. The angular momentum therefore  flips sign 
when seen from different asymptotic regions, as it should be. 
Indeed, if the 
observer as $x\to\infty$   sees the wormhole spin clockwise, say,  
then the observer at $x\to-\infty$  will see it spin in the opposite direction. 

The above discussion implies that the metric amplitude $V$ can be chosen symmetric with respect to reflections 
in the wormhole throat, $x\to-x$, hence 
\be            \label{cond}
\V(-x,y)=\V(x,y)~~~~\Rightarrow~~~
\left.\partial_x \V\right|_{x=0}=0.
\ee
The rotation field $\W$ has different signs in the asymptotic regions, which suggests that it should be antisymmetric and
hence vanish at $x=0$. However, one can show that it cannot vanish both at $x=0$ and at $x=\infty$, 
unless it is zero everywhere.
As a result, the procedure is slightly more involved.

The equations contain only derivatives of the rotation field $\W$,
hence this function
is determined up to an additive constant: if $\W$ is a solution,
then $\underline{\W}=\W+\mathrm{const.}$ is also a solution. Using this freedom,
one can choose $\underline{\W}$ to be antisymmetric,
\be                 \label{47}
\underline{\W}(-x,y)=-\underline{\W}(x,y)~~~~\Rightarrow~~~
\left.\underline{\W}\right|_{x=0}=0. 
\ee
However, it will not vanish in the asymptotic regions and will approach a non-zero value denoted by $-\W_0$ for $x\to\infty$,
therefore approaching $+\W_0$ for $x\to-\infty$:
\be               \label{48}
\W_0+\frac{2J}{x^3}+\ldots ~~~\leftarrow ~~~\underline{\W}\rightarrow ~~~~~ -\W_0+\frac{2J}{x^3}+\ldots ~~~~
\text{as}~~~ -\infty \leftarrow x\rightarrow \infty. 
\ee
At the same time, the line element \eqref{METR} is invariant under 
\be              \label{twis}
\W\to \W+\Omega,~~~~~\varphi\to \varphi+\Omega\,t,
\ee
which amounts to passing to a rotating frame.
Using this symmetry with $\Omega=\pm \W_0$, one can produce from $\underline{\W}$ two different rotation fields, 
$\W_{+}$ and $\W_{-}$,
one of which vanishes asymptotically for $x>0$, and the other for $x<0$:
\be                \label{Wpm}
x>0:~~~~\W_{+}&=&\underline{\W}+\W_0,~~~~~ \W_{+}=\frac{2J}{x^3}+\ldots ~~~ \text{as}~~~ x\rightarrow \infty;\nn \\
x<0:~~~~\W_{-}&=&\underline{\W}-\W_0,~~~~~ \W_{-}=\frac{2J}{x^3}+\ldots ~~~ \text{as}~~~ x\rightarrow -\infty. 
\ee
Therefore, the boundary conditions at infinity are fulfilled at the expense of using two different rotation frames, one for $x>0$ and
the other for $x<0$. One has
\be
\W_\pm = \frac{2J_\pm}{|x|^3}+\ldots~~~~~ \text{as}~~~  |x|\to\infty~~~~\text{where}~~~~J_\pm =\pm J,
\ee
so that the  angular momentum seen  from the two infinities  has different signs. 

Setting   $\W=\W_+$ for $x>0$ and  $\W=\W_-$ for $x<0$  yields 
\be
\lim_{x\to 0}\W=\W_0,~~~~~~\W= \frac{2J}{x^3}+\ldots~~~~~ \text{as}~~~  x\to\infty,
\ee
while for $x<0$ 
\be              \label{cond1}
\W(-x,y)=-\W(x,y).
\ee
The asymptotic conditions are then fulfilled and the rotation field is antisymmetric,
but at the expense of a discontinuity at $x=0$, because one has
\be
\lim_{x\to\pm 0}\W(x,y)=\pm \W_0\,. 
\ee
This is the price of using two different rotation frames. On the other hand,
the value of $\W$ at $x\to 0$ is the throat angular velocity, which  should change sign
depending on the asymptotic region from which the system is observed, since the angular momentum also changes sign.
Therefore, it is natural that the rotation field $\W$ has different limits when approaching the throat from different sides.
In any case, one can always return to the field $\underline{\W}$, which is continuous.

Next, we require the solutions to be symmetric with respect to reflections in the 
the equatorial plane $y=0$, 
\be                \label{cond2}
\V(x,-y)=\V(x,y),~~~~~\W(x,-y)=\W(x,y), 
\ee
hence $\V_{,y}=\W_{,y}=0$ at $y=0$. 

Finally, at the symmetry axis, where $\vartheta=0$ and $y=\cos\vartheta=1$, 
one should impose  the Neumann boundary condition,
\be
\left.\frac{\partial}{\partial\vartheta}\right|_{\vartheta=0}=-\left.\sqrt{1-y^2}\, \frac{\partial}{\partial y}\right|_{y=1}=0,
\ee
hence the derivatives $\V_{,y}$ and $\W_{,y}$ should be finite at $y=1$.

Summarizing the above discussion, to construct the solutions for $-\infty<x<\infty$ and $-1\leq y\leq $ 
it is sufficient to obtain  them within the domain where $x$ and $y$ are positive, 
\be
{\cal D}=\left\{0\leq x<\infty,~0\leq y\leq 1\right\},
\ee
assuming the following boundary  conditions: 
\be                    \label{bc}
\underline{\text{throat},~x=0,~ y\in[0,1]}:&& ~~~\V_{, x}=0,~~~\W=\W_0,~~~~\chioo_{,x}=0;~~~~\nn \\
\underline{\text{infinity},~x=\infty,~ y\in[0,1]}:&& ~~~\V=0,~~~~\W=0,~~~~~~~\chioo_{,x}=0;~~~~\nn \\
\underline{\text{equator},~0\leq x<\infty,~ y=0}:&& ~~~\V_{, y}=0,~~~\W_{, y}=0,~~~~~~~\chioo=0;~~~~\nn \\
\underline{\text{symmetry axis},~0\leq x<\infty,~ y=1}:&& ~~~|\V_{, y}|<\infty,~|\W_{, y}|<\infty,~~\chioo=4J. 
\ee
One then uses the rules \eqref{cond},\eqref{cond1}, and \eqref{cond2} to extend the solutions to negative $x$ and $y$.

We have included in \eqref{bc} also the twist potential $\chioo$ related to $\W$ via \eqref{twisto}, hence 
\be                   \label{420}
\chioo_{,y}=(x^2+\nuu )^2(y^2-1) e^{-4\V}{\W}_{,x},~~~~~
\chioo_{,x}=(x^2+\nuu )(1-y^2)^2 e^{-4\V}{\W}_{,y}. 
\ee
The latter of these relations implies that $\chioo_{,x}=0$  everywhere at the boundary of ${\cal D}$. Indeed, one has $\chioo_{,x}=0$
at $x=0$, where ${\W}_{,y}=0$. This is also true  at $x=\infty$, since $\W\sim1/x^3$ for $x\to\infty$; also at $y=0$, where ${\W}_{,y}=0$;
and finally at  $y=1$.

It follows that  $\chioo(x,y=0)$ is a constant that can be set to zero:  $\chioo(x,y=0)=0$.  
The value $\chioo(x,y=1)$ is also constant, and it can be computed either for $x\to\infty$ or for $x\to 0$. 
Using the first relation in \eqref{420} and the asymptotic form of $\V,\W$ for $x\to\infty$, 
one has 
\be
\lim_{x\to\infty} \chioo(x,y=1)=\lim_{x\to\infty} \int_0^1 \chioo_{,y}(x,y)\, dy
=\lim_{x\to\infty} \int_0^1 (x^2+\nuu )^2(1-y^2) e^{-4\V}{\W}_{,x}\, dy=4J.~~~~
\ee
Therefore, one has $\chioo(x=\infty,y=1)=4J$. 

On the other hand, integrating
the $\W$-equation in \eqref{UUU} over the region ${\cal D}$  and using the boundary conditions \eqref{bc}  yields 
\be
\left.\int_0^1 dy\, \rho^2\, e^{-4\V}(x^2+\nuu )\W_{, x}\right|_{x=0}^{x=\infty}=0\,. 
\ee
Using again the asymptotic form of $\V,\W$ for $x\to\infty$ and 
 the first relation in \eqref{420}, this reduces to 
\be       \label{J}
4J=\left.a^4\int_0^1 dy\,(y^2-1) e^{-4\V} \W_{, x} \right|_{x=0}=\left. \int_0^1 dy\,\chioo_{,y}\, \right|_{x=0}=\chioo(x=0,y=1). 
\ee

As a result, one has  $\chioo=0$  at the equator and  $\chioo=4J$ at the symmetry axis. This explains the boundary 
conditions for $\chioo$  in  \eqref{bc}.

\subsection{Mass, quadrupole  moment, angular momentum}

As discussed in \ref{AppC}, integrating the $\V$-equation in \eqref{UUU} over the domain ${\cal D}$ yields the 
integral representation for the ADM mass, 
\be             \label{M}
M=\frac12\int_{\cal D} \,\rho^2 e^{-4\V}\langle \W,\W\rangle \, dx\,dy\, , 
\ee
and a similar  representation for the  quadrupole moment $Q$, 
\be             \label{Q}
Q=-\frac14\int_{\cal D} \,(2z^2-\rho^2) e^{-4\V}\langle \W,\W\rangle \, dx\,dy+\frac{M^3}{3}. 
\ee

A similar method can be used to relate the mass $M$ and angular moment $J$ \cite{Kleihaus:2014dla}. 
The two Ernst equations in \eqref{EEEx} can be combined into 
\be
\nabla_k\nabla^k \V=\nabla_k
\left(\frac{\rho^2}{2}\,e^{-4\V}\W\nabla^k\W\right),
\ee
or explicitly 
\be                     
\label{UUUab}
[(x^2+\nuu )\V_{,x}]_{,x}+[(1-y^2)\V_{,y}]_{,y}
=
\left[\frac{\rho^2}{2}\,e^{-4\V}\,(x^2+\nuu )\,\W\W_{,x}\right]_{,x}+\left[\frac{\rho^2}{2}\,e^{-4\V}\, (1-y^2)\, \W\W_{,y}\right]_{,y}.~~~~
\ee
Integrating this  over ${\cal D}$, using  the asymptotic form of $\V,\W$ for $x\to\infty$,  the boundary conditions \eqref{bc},  
and the first relation in \eqref{420}, yields 
\be
\left. \int_{0}^{1} dy\, (x^2+\nuu )\, \partial_x \V\right|_{x=0}^{x=\infty}&=&
\frac12 \left. \int_{0}^{1} dy\,(1-y^2) (x^2+\nuu )^2\,e^{-4\V} \W\partial_x \W \right|_{x=0}^{x=\infty} \nn \\
&=&\left.\frac12\, W\right|_{x=0}\left.\int_0^1 dy\, \chioo_{,y}\right|_{x=0}, 
\ee
therefore one obtains  \cite{Kleihaus:2014dla,Chew:2016epf} 
\be             \label{MJ}
M=2\,\W_0 J.
\ee
This is similar to the Smarr relation for black holes \cite{Smarr:1972kt}.
This shows, in particular, that if $\W$ vanishes at the throat, it cannot vanish at infinity.
This also shows that both $\W_0$ and $J$ should change sign 
when viewed from different sides of the throat, since $M$ does not change sign.
The values of $J$ and $M$ obtained from \eqref{J} and \eqref{M} should fulfill the relation \eqref{MJ},
which provides a good check of the numerical procedure.

The boundary conditions \eqref{bc} are invariant under 
\be                   \label{vir}
\V\to \lambda\, \V,~~~~\W\to \W,~~~~\chioo\to \chioo,
\ee
with a constant $\lambda$. Therefore, the functionals $E_{\W}$ and $E_{\chioo}$ in \eqref{Ew},\eqref{Ec} 
should be stationary with respect to this transformation at $\lambda=1$. This yields the virial 
relations for solutions of the Ernst equations, 
\be                 \label{vir1}
dE_{\W}&\equiv& \int_{\cal D}\left( \langle \V,\V\rangle+\frac{\rho^2 \V}{2}\,e^{-4\V}\langle \W,\W\rangle \right) dx \,dy =0, \nn \\
dE_{\chioo}&\equiv& \int_{\cal D}\left( \langle \V,\V\rangle+\frac{\V}{2\rho^4}\,e^{4\V}\langle \chioo,\chioo\rangle \right) dx \,dy =0.
\ee
These relations also provide a good check of the consistency of the numerical procedure.

Finally, it is worth noting that, when passing to the dimensionful metric  $d{\rm s}^2=\upmu^2 ds^2$, 
one should replace 
 $M,J,Q$ by their
dimensionful versions, 
\be                 \label{scale}
{\rm M}=\upmu M,~~~~~~{\rm J}=\upmu^2 J,~~~~~~~{\rm Q}=\upmu^3 Q.
\ee

\section{Numerical solutions \label{SctVI}}
\setcounter{equation}{0}

%We are now ready to solve the Ernst equations with the boundary conditions \eqref{bc}. 
%\blue{Within the numerical approach, this is not a very challenging problem that can be analyzed by various  methods. }
We are now ready to solve the Ernst equations subject to the boundary conditions \eqref{bc}.
\blue{
Numerically, this is a relatively straightforward problem that can be approached using several different methods.}
We employ  the
FreeFem++ numerical solver based on the finite element method \cite{MR3043640}.
This solver uses the weak form of the equations defined by the functional $E_{\W}$ in \eqref{Ew},
or by $E_{\omega}$ in \eqref{Ec},
and by their first and second
variations. The variations of $\V,\W$ are expanded
with respect to basis functions obtained by triangulating the integration domain ${\cal D}$,
while
the non-linearities are handled within the Newton--Raphson scheme.

We compactify the integration domain by using, instead of $x\in[0,\infty)$, the variable
\be
{\rm x}=\frac{x}{1+x}\in[0,1),
\ee
and we set the throat size parameter to unity,
\be
a =1.
\ee
As explained below, solutions for other values of $a$ can be obtained from the $a=1$ solutions
via a simple rescaling.

The boundary conditions in \eqref{bc} depend on one parameter, which is the throat angular velocity $\W_0$ for the
$(\V,\W)$-equations, or the angular momentum $J$ for the $(\V,\chioo)$-equations. We first consider the $(\V,\W)$-equations
and iteratively increase the value of $\W_0$, starting from $\W_0=0$. The numerical procedure converges
to a smooth solution for $\V,\W$.
The convergence is controlled by the virial relation $dE_{\W}=0$ in \eqref{vir1},
which is fulfilled with a precision depending on the numbers of the discretization points $N_{\rm x}$
and $N_y$ along the ${\rm x},y$ axes (these numbers determine the triangulation pattern for the FreeFem++ solver).
Taking $N_{\rm x}=N_y=100$ typically yields
$dE_\W\sim 10^{-15}$. Having constructed $\V,\W$, we integrate the first-order equations \eqref{KKr} and \eqref{420}
to obtain $\K_{\rm reg}$ and $\chioo$.

\blue{
As an additional check, we compare our results with the analysis of the dressed wormholes reported previously in 
Refs.~\cite{Kleihaus:2014dla,Chew:2016epf}, and successfully reproduce their results within our numerical approach.
}

\begin{figure}[th]
\hbox to \linewidth{ \hss

	\includegraphics[width=8 cm]{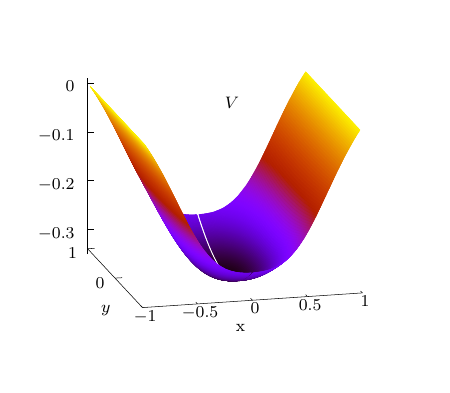}
	\includegraphics[width=8 cm]{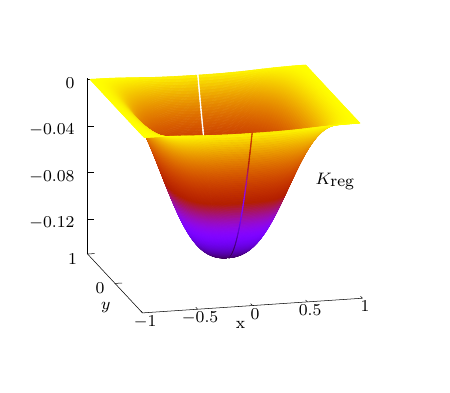}	
	
\hss}
\hbox to \linewidth{ \hss

	\includegraphics[width=8 cm]{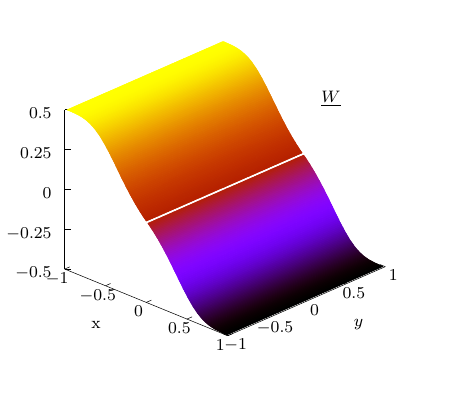}
	\includegraphics[width=8 cm]{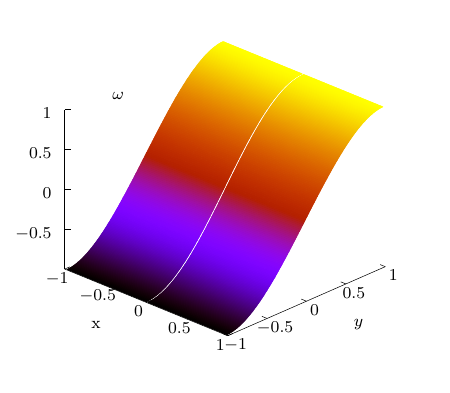}	
\hss}
\caption{Profiles of $\V,\underline{\W},\K_{\rm reg},\chioo$ against the 
compactified radial coordinate ${\rm x}\in (-1,1)$ and $y\in [-1,1]$ for $\W_0=1/2$.
The central cuts in the plots correspond to the position of the throat.}
\label{Fig1}
\end{figure}

Solutions obtained in the domain ${\cal D}$  can be extended to the whole space  $x\in(-\infty,\infty)$ and $y\in[-1,1]$ 
using the reflections $x\to -x$ and $y\to -y$. 
The corresponding profiles of $\V,\underline{\W},\K_{\rm reg}$, and $\chioo$  are shown in Fig.\ref{Fig1} for $\W_0=1/2$; 
for other values of $\W_0$ they are 
qualitatively similar. The amplitude $\V$ approaches  zero at infinity  and achieves its minimum  value 
in the equatorial part of the throat, which corresponds to the position of the ring, 
\be               \label{V0}
\V_0\equiv \min_{x,y} \V(x,y)=\V(0,0)<0.
\ee
This determines the ring radius $R_{0}$ and its 
 linear  velocity $v_{0}$, as well as the ring tension   $\C$ defined in  \eqref{Kr2o}, 
\be                        \label{C}
R_{0}=e^{-\V_0},~~~~~~~~~v_0=\W_{0} R_0\,,~~~~~~~\C=1-\frac{R_0^4}{4}\,\W^2_{,x}(0,0). 
\ee
The metric amplitude ${\K}_{\rm reg}$  also shows an absolute minimum 
at the ring, 
\be
\K_0\equiv \min_{x,y} {\K}_{\rm reg}(x,y)={\K}_{\rm reg}(0,0)<0,
\ee
and it vanishes on the symmetry axis as it should, ${\K}_{\rm reg}(x,y=1)=0$. 
The amplitudes $\V$ and $\K_{\rm reg}$ are symmetric under $x\to -x$ and $y\to -y$.

As discussed above, the continuous and globally defined rotation field $\underline{\W}$ is antisymmetric under $x\to -x$, 
but does not vanish asymptotically. Using two local rotation frames, it  
determines two local fields $\W_\pm =\underline{\W}\pm \W_0$ which approach zero,
respectively, for $x\to\pm \infty$. 
The twist field $\chioo$  is antisymmetric under $y\to -y$  and 
assumes the constant value $\chioo=4J$ at the symmetry axis, where $y=1$.

The rotation field $\underline{\W}(x,y)$ shows a very weak dependence on $y$, and its level lines are approximately
parallel to the $y$-axis. Similarly, the twist field $\chioo(x,y)$ has a very weak dependence on $x$, and its level lines are approximately
parallel to the $x$-axis. The lines of constant $\underline{\W}$ and constant $\chioo$ are mutually orthogonal.

\begin{figure}[t]
\hbox to \linewidth{ \hss
	
	\includegraphics[width=12 cm]{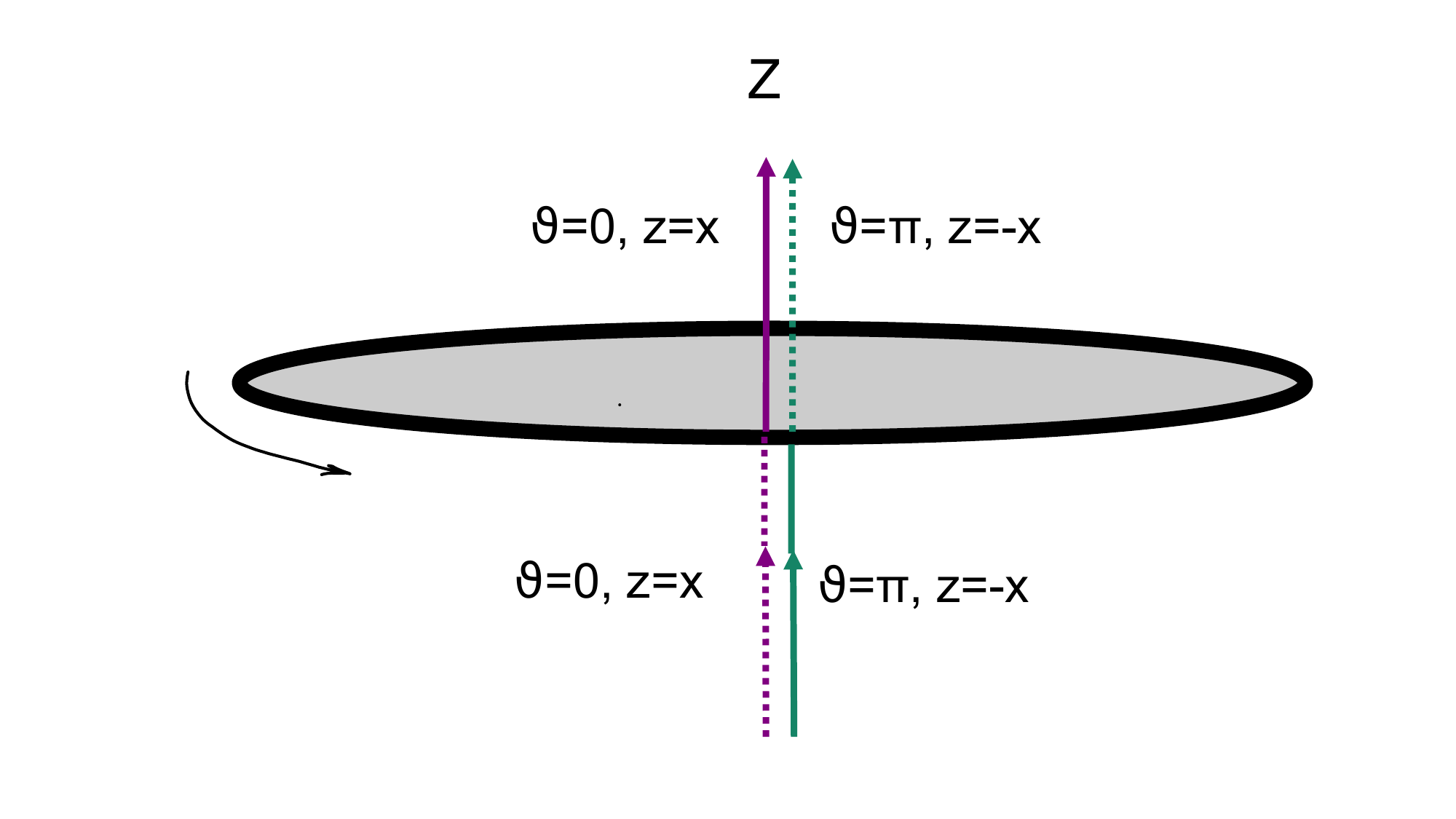}

\hss}
\caption{
The spinning wormhole has two rotation axes that extend from one asymptotic region to the other:
$\vartheta=0$, with $z=x\in(-\infty,\infty)$, and
$\vartheta=\pi$, with $z=-x\in(-\infty,\infty)$.
The solid lines correspond to the portions of the axes belonging to the $x>0$ region.
}
\label{Fig1oo}
\end{figure}

The ADM mass $M$ and angular momentum $J$ are obtained by integrating \eqref{M} and \eqref{J},
and we verify that the relation $M=2\W_0 J$ indeed holds. For slow rotation with $\W_0\ll 1$, one recovers
the perturbative result \cite{Volkov:2021blw},
\be \label{slow}
M=\frac{4}{3\pi}\, \W_0^2+\ldots,~~~~~~~J=\frac{2}{3\pi}\,\W_0+\ldots.
\ee
One may compare this with the non-relativistic expressions for rotating rigid bodies,
\be \label{Erot}
E_{\rm rot}\equiv M=\frac{J^2}{2I}.
\ee
This formula agrees with the field-theoretic result \eqref{slow}
provided that the moment of inertia is $I=1/(6\pi)$, but it is
not clear how to interpret this value of $I$.

Moreover, inserting it into the angular momentum formula 
$
J=I\,\dot{\theta}
$
yields the angular velocity
$\dot{\theta}=4\W_0$ rather than $\dot{\theta}=\W_0$, as one might naively expect.
One may think  that the additional factor of four arises because the rotation occurs 
simultaneously in the two spaces corresponding to $x>0$ and $x<0$,
and also around the two rotation axes corresponding to $x\in(-\infty,\infty)$ 
and either $y=\cos\vartheta=1$ or $y=-1$ (see Fig.\ref{Fig1oo}).
However, the analogy with a rotating rigid body may be incomplete, 
since the spinning wormhole is ultimately a field configuration rather than a solid.

\begin{figure}[th]
\hbox to \linewidth{ \hss

	\includegraphics[width=8 cm]{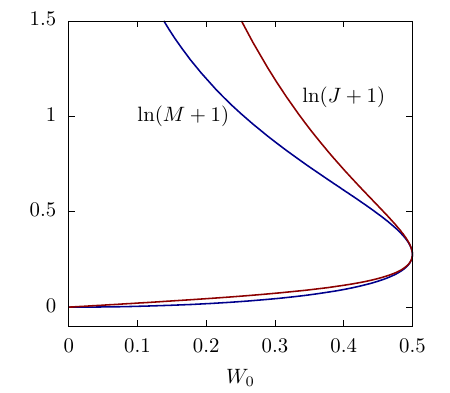}	
		\includegraphics[width=8 cm]{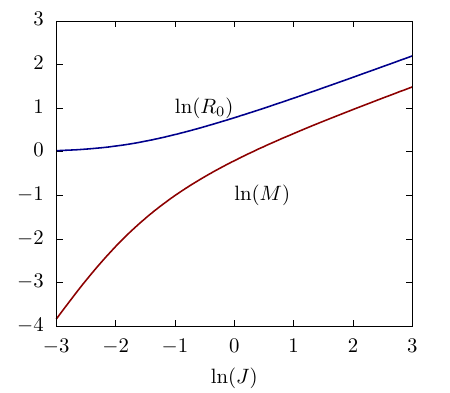}

\hss}

\caption{Left: the ADM mass $M$ and angular momentum $J$ against the throat angular velocity $\W_0$. 
For each $\W_0\in (0,1/2)$ there are two solutions: one slowly rotating and  one rapidly rotating. 
Right: the throat radius  $R_0$ and mass $M$ against the angular momentum $J$. 
}
\label{Fig2}
\end{figure}

When $\W_0$ increases, both $M$ and $J$ grow. However, $\W_0$ increases 
only up to the maximal value $\W_0=1/2$ and then starts decreasing,
while $M$ and $J$ continue to grow, as shown in Fig.\ref{Fig2}. In other words, for each given $\W_0\in (0,1/2)$
there are two spinning solutions: one belonging to the “slow rotation branch” 
with $M<M_\otimes$, $J<J_\otimes$, and the other
belonging to the “fast rotation branch” with $M>M_\otimes$, $J>J_\otimes$. \blue{The two branches merge
when the mass and angular momentum coincide, with their common values given by 
\be \label{crit}
M_\otimes=J_\otimes=0.318\ldots \approx \frac{1}{\pi}.
\ee
Equation \eqref{MJ} then implies that $M/J=2\W_0=1$, so that  the merging occurs  precisely  at 
$\W_0=1/2$. 
}
To recover both solution branches, it is more convenient to use the $(\V,\chioo)$ equations. 
In this case, the value of $J$ is prescribed,
while the rotation field $\W$ is obtained from \eqref{twisto}.

\begin{figure}[th]
\hbox to \linewidth{ \hss

\includegraphics[width=8 cm]{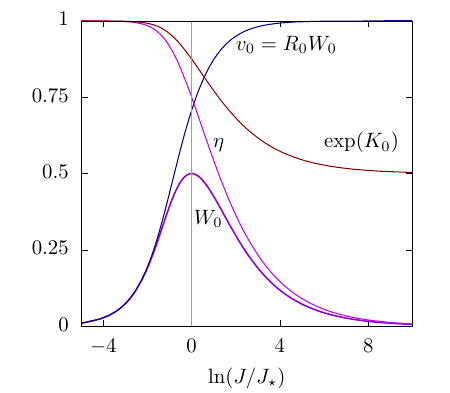}	
\includegraphics[width=8 cm]{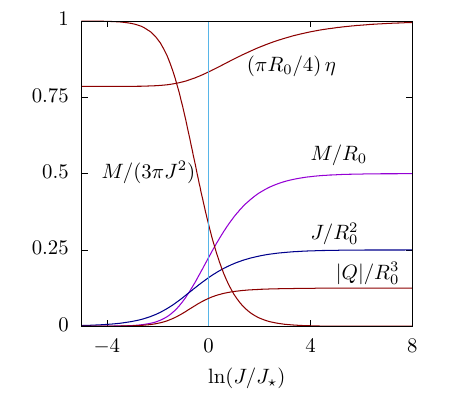}	

\hss}

\caption{Parameters of spinning solutions. }
\label{Fig3}
\end{figure}

Various parameters against the angular momentum $J$ are
shown in Fig.\ref{Fig2} and Fig.\ref{Fig3}.
When $J$ increases, the mass $M$ and the throat radius $R_0$ grow without bounds,
\be
0\leftarrow M\to \infty,~~~~1\leftarrow R_0\to \infty~~~~~~\text{as}~~~~~0\leftarrow J\to \infty.
\ee
The other parameters are bounded. The throat angular velocity $\W_0$ 
increases up to the maximal value $\W_0=1/2$, but then
decreases and approaches zero in the fast-rotation limit. The throat linear velocity $v_0=R_0\W_0$
grows monotonically,
\be
0\leftarrow \W_0\to 0,~~~~~0\leftarrow v_0=R_0\W_0\to 1, ~~~~~\text{as}~~~~~0\leftarrow J\to \infty.
\ee
The tension parameter $\C$ tends to zero for fast rotation, while ${K}_0$ (the minimal value of $\K_{\rm reg}$) 
approaches a finite value,
\be
1\leftarrow \C\to 0,~~~~~1\leftarrow e^{K_0}\to 1/2 ~~~~~\text{as}~~~~~0\leftarrow J\to \infty.
\ee
The ratios $M/R_0$, $J/R^2_0$, $Q/R^3_0$, 
and $\C R_0 $ also approach finite values in the fast rotation limit, 
\be          \label{fast0}
&&0\leftarrow \frac{M}{R_0}\to \frac12,~~~~~
0\leftarrow \frac{J}{R_0^2}\to \frac14,~~~~~
0\leftarrow \frac{|Q|}{R_0^3}\to \frac18,~~~~~
1\leftarrow \C R_0\to \frac{4}{\pi}.~~~~~
\ee
It follows that in the fast rotation limit one has $\W_0\ll 1$ and 
\be             \label{fast}
M=\frac{1}{2\W_0}+\ldots,~~~~~~~J=\frac{1}{4\W_0^2}+\ldots .
\ee
The vertical line in the two panels in Fig.\ref{Fig3} separates the slow-rotation branch and the fast-rotation branch.
The physical difference between them is the position of the ergoregion.
The latter is defined as follows. The rotation field
$
\W=\underline{\W}+\W_{0}
$
shows the asymptotics $2\W_0\leftarrow \W\to {\cal O}(1/r^3)$
as $-\infty\leftarrow x\to \infty$, hence the metric component
\be
g_{00}=-e^{2\V}+\rho^2 \W^2 e^{-2\V}
\ee
fulfills
\be
+\infty \leftarrow g_{00}\to -1~~~~\text{as}~~~-\infty\leftarrow x\to \infty.
\ee
The region where $g_{00}>0$ is the ergoregion, and its boundary, where $g_{00}$ changes sign, is the ergosurface.

\begin{figure}[th]
\hbox to \linewidth{ \hss

\includegraphics[width=8 cm]{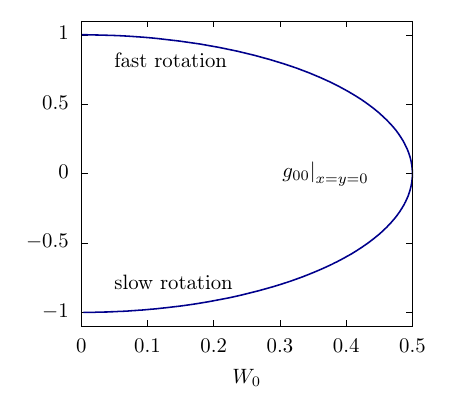}	
\includegraphics[width=8 cm]{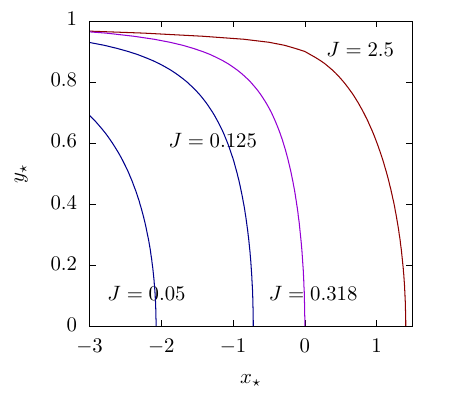}

\hss}

\caption{The value of $g_{00}$ at the ring, where $x=y=0$ (left), and 
the boundary of the ergoregion for different $J$ (right).}
\label{Fig5}
\end{figure}

Let us compute $g_{00}$ in the equatorial part of the throat, at the ring position,
\be
\left.g_{00}\right|_{x=y=0}=-e^{2\V_0}+\W_0^2 e^{-2\V_0} =-\frac{1}{R_0^2}+\W_0^2 R_0^2\,.
\ee
Its numerical plot in Fig.\ref{Fig5} (left panel) shows that it is negative for slow rotation and becomes positive for fast rotation.
The two branches meet at $\W_0=1/2$.

As seen in Fig.\ref{Fig5}, the plot of $g_{00}(\W_0)$ is exactly invariant under a flip of the sign.
This implies that the radius $R_0^{\rm slow}$ of the slowly rotating solution is related
to the radius $R_0^{\rm fast}$ of the rapidly rotating solution as follows:
\be
g_{00}^{\rm slow}(\W_0)=- g_{00}^{\rm fast}(\W_0)\Rightarrow
-\frac{1}{(R^{\rm slow}_0)^2}+\W_0^2 (R^{\rm slow}_0)^2 = \frac{1}{(R^{\rm fast}_0)^2}-\W_0^2 (R^{\rm fast}_0)^2,
\ee
from where
\be \label{dual0}
\W_0R^{\rm slow}_0R^{\rm fast}_0=1.
\ee

Away from the equatorial plane, the position of the ergosurface is 
determined by the coordinate values $x_\star,y_\star$ for which $g_{00}=0$, hence
\be
e^{2\V(x_\star,y_\star)}=\sqrt{(x_\star^2+1 )(1-y_\star^2)}\,\W(x_\star,y_\star).
\ee
The solution of this equation for various $J$ is shown in Fig.\ref{Fig5}. 
The ergosurface extends to $x_\ast\to -\infty$ when $y_\ast\to 1$, approaching the symmetry axis. The coordinate $x_\star$ reaches its maximal value at the equator, where $y_\star=0$.
The values of $x_\star$ for $y_\star=0$ fit the formula
\be               \label{xstar}
\sqrt{x_\star^2+1}=\frac{1}{2\W_0}~~~~\Rightarrow~~~~x_\star=\pm \frac{\sqrt{1-4\W_0^2}}{2\W_0}\,.
\ee
Here the minus sign corresponds to the slow branch, for which the ergosurface lies entirely in the $x<0$ region. 
The plus sign corresponds to the fast branch, where part of the ergosurface intersects the wormhole 
throat and extends into the $x>0$ region. For $\W_0=1/2$ the ergosurface touches the throat at the ring position.

\blue{
One should emphasize that we do not have an analytic explanation for the relation \eqref{xstar}, 
although it is satisfied numerically to high precision. A similar relation, expressed in different coordinates, 
was reported in Refs.~\cite{Kleihaus:2014dla,Chew:2016epf}.
}

Another  relation that, similarly to \eqref{dual0},
expresses  a duality between the slow rotation and fast rotation can be obtained using the formula $M=2\W_0 J$ 
in  \eqref{MJ}. Applying it separately to the  slow rotation and fast rotation yields 
\be
M_{\rm slow}=2\W_0 J_{\rm slow},~~~~~~M_{\rm fast}=2\W_0 J_{\rm fast},
\ee  
therefore 
\be
2\W_0=\frac{M_{\rm slow}}{J_{\rm slow}}=\frac{M_{\rm fast}}{J_{\rm fast}}\,,
\ee
hence 
\be
M_{\rm slow} J_{\rm fast}=
M_{\rm fast}J_{\rm slow} =2\W_0 J_{\rm slow} J_{\rm fast} \equiv f^2(\W_0)\,.
\ee
It turns out that the function $f(\W_0)$ is almost constant, since Eqs.\eqref{slow},\eqref{fast} imply that 
for $\W_0\to 0$ it approaches the value 
\be
f(0)=\frac{1}{\sqrt{3\pi}}=0.325\ldots ,
\ee
while for  $\W_0\to 1/2$ one obtains from \eqref{crit} 
\be
f(1/2)=M_\otimes=J_\otimes=0.318\ldots ,
\ee
the relative difference between  these two values being about  $2\%$.

It follows that 
the above numerical plots are approximately invariant under 
\be
M\to f^2/J,~~ ~~~~~J\to {f}^2/M,
\ee
where $f$ is a constant in the interval $[0.318,0.325]$. 

\blue{
It is also worth noting that, although $g_{00}$ changes sign, $g_{\varphi\varphi}=\rho^2 e^{-2\V}$ remains everywhere non-negative, so there are no closed timelike curves.
}

\section{Scaling \label{SctVII}}
\setcounter{equation}{0}

As the wormhole spins faster, the ring radius $R_0$ increases because the ring is stretched by the centrifugal force. 
According to Eq.\eqref{dual0}, one has
$R_0^{\rm fast}=1/(\W_0 R_0^{\rm slow})$,
hence in the fast rotation limit, when $\W_0\to 0$ and $R_0^{\rm slow}\to 1$, the radius $R_0=R_0^{\rm fast}\to\infty$.

The other parameters of the solutions also grow with $R_0$. 
However, as seen in Fig.\ref{Fig3} and shown by Eq.\eqref{fast0}, 
rescaling the parameters by powers of $R_0$ yields finite values in the fast rotation limit. 
It turns out that such a rescaling can be incorporated into the solutions from the very beginning.

Let us remember that the above solutions were obtained for the fixed size parameter,
$a =1$, which corresponds to the unit ring radius in the static limit. They are characterized by
\be                   \label{nR}
&&a =1,~~~~R_0(J)=a\, e^{-V_0}\,\in [1,\infty),~~~~~\nn \\
&&M(J)\in [0,\infty),~~~~
J\in [0,\infty),~~~~~v_0(J)=R_0 W_0 \in [0,1). 
\ee
These solutions can be generalized for any $a\in (0,\infty)$ by noting that, if $a$ is not fixed, 
the field equations \eqref{UUU},\eqref{UUU1} admit a scaling symmetry. 
Specifically, if there is a solution described by $a$, $V(x,y)$, $\W(x,y)$, $\chioo(x,y)$ with parameters $M,J,Q$, then setting
\be         \label{scale0}
\tilde{a}=\frac{a }{\lambda},~~~\tilde{\V}(x,y)= \V({x}/{\lambda},y),~~~\tilde{\W}(x,y)= \lambda\,\W({x}/{\lambda},y),
~~~\tilde{\chioo}(x,y)= \frac{1}{\lambda^2}\,\chioo({x}/{\lambda},y),
\ee
\purple{for any positive real} $\lambda$, yields another solution with parameters
\be              \label{scale1} 
\tilde{R}_0=\frac{R_0}{\lambda},~~~ 
\tilde{M}=\frac{M}{\lambda},~~~ 
\tilde{J}=\frac{J}{\lambda^2},~~~ 
\tilde{Q}=\frac{Q}{\lambda^3}.
\ee
Applying this symmetry to the $a=1$ solutions \eqref{nR} yields solutions with $a \neq 1$.

The scaling does not change the linear velocity of the ring and its tension,
\be
0\leftarrow v_0=\W_0 R_0\to 1,~~~~~~~~1\leftarrow\C=1-\left.\frac{R_0^4}{4}\,\W_{,x}^2\right|_{x=y=0}\to 0,
\ee
other scale-invariant quantities being
\be                      \label{7.5}
0\leftarrow \frac{M}{R_0}\rightarrow \frac12,~~~~~~~~
0\leftarrow \frac{J}{R_0^2}\rightarrow \frac14,~~~~~~~~
0\leftarrow \frac{M^2}{J}\rightarrow 1,~~~~~~~~0\leftarrow \frac{MQ}{J^2}\rightarrow 1,
\ee
where  the arrows indicate the values in the slow/fast rotation limits $0\leftarrow v_0\rightarrow 1$
(see \eqref{fast0}).

Note that in the fast rotation limit one obtains the Regge relation
$J=M^2$.

As a result, spinning solutions actually comprise a two-parameter set whose members
can be labeled by $J$ and by $a$. Since $J$ changes under scalings, it is sometimes
convenient to use instead the scale-invariant linear velocity $v_0\in [0,1)$ to label the solutions.

When $a$ is fixed, the mass $M$, angular momentum $J$, and radius $R_0$ of the solutions grow without bounds
with increasing rotation as $v_0\to 1$.
This is natural, since fixing $a$ determines the size of the ring in the static limit, and as it spins faster,
it is stretched by the centrifugal force, which increases its energy and angular momentum.

Notice, however, that the scaling parameter $\lambda$ in \eqref{scale0},\eqref{scale1} does not need
to be always constant and may depend on the rotation, $\lambda=\lambda(v_0)$.
This can be used to obtain spinning solutions whose radius $R_0$ does not change with the rotation.
Let us choose in \eqref{scale0}, \eqref{scale1}
\be 
\lambda(v_0)=\frac{R_0(v_0)}{2\m}, 
\ee
where $\m$ is a constant. Both $R_0(v_0)$ and $\lambda(v_0)$ grow when $v_0$ increases. 
However, inserting $\lambda(v_0)$ into \eqref{scale0},\eqref{scale1} 
and omitting the tilde sign yields solutions whose parameters are always finite,
\be                    \label{Rn}
&&a=2\m e^{V_0}\in (0,2\m],~~~~~
R_0=2\m,~~~~~\nn \\
&&M\in [0,\m),~~~~
J\in [0,\m^2),~~~~
|Q|\in [0,\m^3). 
\ee
Such solutions can also be constructed by directly solving Eqs.\eqref{UUU},\eqref{UUU1} 
and adjusting the value of $v_0$ to fulfil the condition $R_0(a,v_0)=2\m$, which determines $v_0=v_0(a)$.
\blue{One should emphasize that ${\cal M}$ is not the mass but the parameter that fixes the ring radius $R_0$, 
although it determines also the maximal value of the mass $M$. }

One can wonder why, for these solutions, the ring radius $R_0$ is always the same and not stretched by the 
centrifugal force, and why $M$ and $J$ remain bounded. The answer is that these solutions describe 
not one spinning ring but a family of rings of different size.

Specifically, consider a family of {\it static rings} of sizes $a\in(0,2\m]$. If they start spinning, they 
become stretched. When the ring of size $a$ spins with the linear velocity $v_0(a)$, it stretches 
up to the radius $R_0=2\m$. Applying this for all values of 
$a$, all rotating rings will stretch to the same radius, which corresponds to the solutions in \eqref{Rn}.

If $a$ is only slightly below $2\m$, then a slow rotation is sufficient to stretch the ring to the size $R_0=2\m$, 
but if $a$ is close to zero, then a fast rotation is needed to reach the same size. 
The fast rotation limit corresponds to a spinning ring which is vanishingly small in the static limit, 
with the size parameter $a$ approaching zero, but whose linear velocity $v_0$ approaches 
unity in such a way that it is stretched by the rotation only up to a finite size. 
This is why its mass and angular momentum remain bounded.

\section{Fast rotation limit \label{SctVIII}}
\setcounter{equation}{0}

A ring with a fixed size parameter $a$ is stretched to an infinite physical radius $R_0$ 
in the fast-rotation limit $v_0\to1$, while its mass and angular momentum $M,J$ grow without bound.
 
 However, for the fixed radius solutions \eqref{Rn} all parameters remain finite: $R_0=2{\cal M}$, $M<{\mathcal M}$, etc. 
 Their values resemble those for the Kerr metric, which has a fixed equatorial size of the 
 horizon $R_0=2M$ that does not depend on the angular momentum $J\in[0,M^2]$ (see \eqref{A18}). 
 Moreover, in the $v_0\to 1$ limit Eq.\eqref{Rn} yields
\be             \label{wrmM}
M\to \m,~~~~
J\to \m^2,~~~~
|Q|\to \m^3, 
\ee
which is the same pattern as for the extremal Kerr solution (see Eq.\eqref{JM2}).
In addition, one has in the limit
\be
a=2\m e^{\V_0} \to 0. 
\ee
This also resembles the extremal Kerr solution, for which $\nu\equiv a^2\to 0$ (see \ref{AppB}).

\begin{figure}[th]
\hbox to \linewidth{ \hss

\includegraphics[width=8 cm]{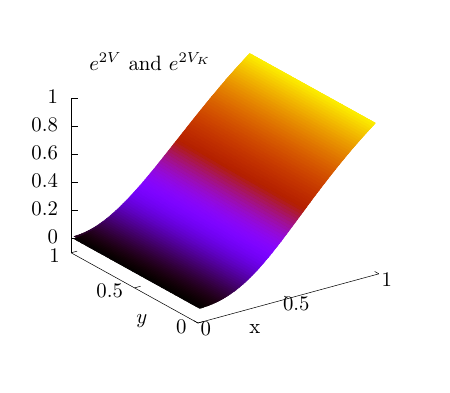}	
\includegraphics[width=8 cm]{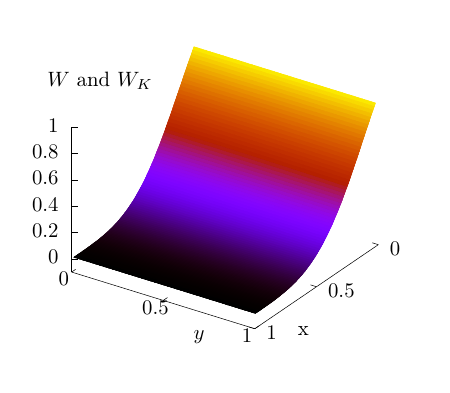}	

\hss}

\caption{Plots of $e^{2V}$ and $W$ for the spinning wormhole with $a^2 =0.01$ and $R_0=1$,
and $e^{2V_K}$ and $W_K$ for the extremal Kerr solution with the equatorial horizon size $R_0=1$.
The profiles of the two solutions almost coincide.
The wormhole throat and the black hole horizon are located at $x={\rm x}=0$.}
\label{Fig6}
\end{figure}

For wormholes, the throat value of $e^\V$ approaches zero in the limit, 
while for the extreme Kerr solution the horizon value of $e^\V$ is exactly zero.
Denoting the amplitudes for the extremal Kerr solution by $\V_{\rm K},\W_{\rm K}$, 
one has (see Eq.\eqref{Kerr33} in \ref{AppA})
\be               \label{Kerr3}
e^{2\V_{\rm K}}=\frac{p^2x^2\,[(px+1)^2+y^2 ] }{\cal Q }, ~~~~
\W_{\rm K}=\frac{2p\,(px+1) }{\cal Q }, 
\ee
where ${\cal Q}=4+px\,[8+px\,(7+px\,(px+4)+y^2 ) ]$ and $p=1/M$ is the inverse black hole mass.
One has for $0\leftarrow x\to\infty$
\be       \label{JM2}
0\leftarrow   e^{2\V_{\rm K}}\to 1-\frac{2M}{x}+\ldots ,~~
\frac{1}{2M}\leftarrow   \W_{\rm K}\to \frac{2 M^2}{x^3}+\ldots \equiv  \frac{2J}{x^3}+\ldots~~\Rightarrow~~J=M^2.~~
\ee
The point $x=0$ in \eqref{Kerr3}  corresponds to the degenerate 
event horizon, where $e^{2\V_K}\sim x^2$, and one has 
 \be
 \left.e^{2\V_{\rm K}}\right|_{x=0}=0, ~~~~~~
 \left.\partial_x(e^{2\V_{\rm K}})\right|_{x=0}=0, ~~~~~~\left.\partial_y \W_{\rm K}\right|_{x=0}=0.
 \ee 
 For wormholes, the point $x=0$ corresponds to the throat, where 
 \be                     \label{8.5}
 \left.\partial_x(e^{2\V})\right|_{x=0}=0, ~~~~\left.\partial_y \W\right|_{x=0}=0.
 \ee 
 \blue{
  The wormholes with fixed $R_0$ therefore satisfy, in the $a\to0$ limit, the same field equations 
 as the extremal Kerr geometry, and the boundary conditions are also similar. 
 This suggests that  the limiting wormhole geometry and the extremal 
 Kerr geometry coincide. However,  this is not guaranteed, since the boundary-value problem is degenerate and the 
 boundary conditions are not exactly the same. The theorem establishing the uniqueness of 
 the Kerr black bole with a degenerate horizon  \cite{Amsel:2009et}
 does not directly apply for wormholes, for which only the derivative 
 $\partial_x(e^{2\V})$ is required to vanish, whereas for the black hole both $e^{2\V}$ itself and its derivative should 
 be zero. 
 }

 Nevertheless, numerical solutions provides evidence  that the spinning wormholes indeed approach 
 the extremal  Kerr geometry as $a\to 0$. 
Fig.\ref{Fig6} shows the profiles of $e^{2V}$ and $\W$
for the wormhole with $a =0.1$ and $R_0=1$, and also $e^{2V_K}$ and $\W_K$ 
for the extremal Kerr black hole with
 the same equatorial horizon size $R_0=1$ (see \eqref{A18}).
The profiles of the two solutions are almost identical, the main difference between 
them being that for $x\to 0$ one has
\be
e^{2V_{\rm K}}\sim x^2,~~~~~e^{2V}\sim x^2+\nuu .
\ee
%This suggests that an analytic approximation for the ``near-extremal'' 
%wormholes with $a \ll 1$ can be obtained from
%\eqref{Kerr3} by replacing $x^2\to x^2+\nuu $ in the expression for $e^{2V_K}$.
In the $a\to 0$ limit, the proper length of the wormhole throat approaches infinity, 
and the throat transforms into a degenerate event horizon.

\begin{figure}[th]
\hbox to \linewidth{ \hss

\includegraphics[width=8 cm]{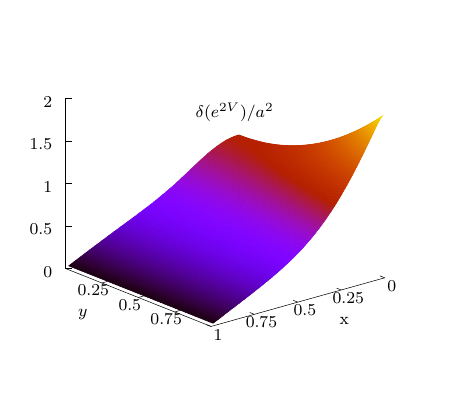}	
\includegraphics[width=8 cm]{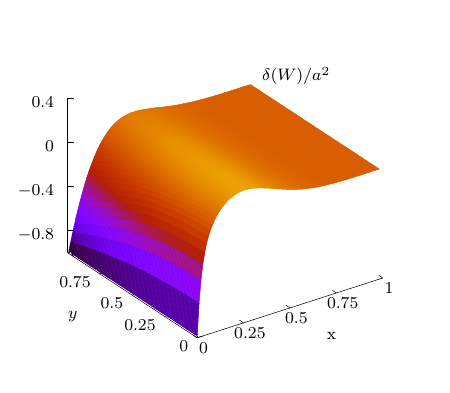}

\hss}

\caption{\blue{The rescaled differences   $\delta(e^{2V})/a^2=\left(e^{2\V}-e^{2\V_{\rm K}}\right)/a^2$ and 
$\delta(\W)/a^2=\left( \W-\W_{\rm K}\right)/a^2$ 
between the profiles of the wormhole with $a^2 =10^{-3}$ and $R_0=1$ and those of 
the extremal Kerr black hole  with
 the equatorial horizon radius $R_0=1$ (see \eqref{A18}). Decreasing  $a^2$ 
 leaves these plots unchanged within numerical accuracy. 
 }
}
\label{Fig66}
\end{figure}

\blue{Fig.\ref{Fig66} shows the profiles of the deviations  $\delta(e^{2V})=e^{2\V}-e^{2\V_{\rm K}}$  
and $\delta(\W)=\W-\W_{\rm K}$  
from the 
extremal Kerr black geometry. The deviations scale as $a^2$, and their norm, 
defined as 
\be                    \label{delta}
\delta(a)=\int_{\cal D}\left(\left| \delta\left(\e^{2V}\right) \right|+\left| \delta(\W) \right|\right) dx\, dy,
\ee
decreases as $a^2$. The rescaled deviations $\delta(e^{2V})/a^2$, $\delta(\W)/a^2$, as well as 
 the rescaled norm $\delta(a)/a^2$, approach finite limits as $a\to 0$. 
For example, for solutions with fixed radius $R_0=1$, we find 
$\delta(a)/a^2=0.10;$ $\,0.13;$ $0.14$   for $a^2=10^{-3}; 10^{-4}; 10^{-5}$, respectively.
\\
\indent
One can understand why the deviations  scale as $a^2$. 
The limiting Kerr configuration is described by $e^{2\V_{\rm K}},\W_{\rm K}$, 
which  satisfy  the Ernst equations with $a^2=0$. 
The  rapidly spinning wormhole is described by 
$e^{2V}=e^{2K_{\rm K}}+\delta\left(e^{2\V} \right) $ and $\W=\W_{\rm K}+\delta\left(\W \right)$, 
which satisfy  the Ernst equations with $a^2\ll 1$. 
Since both the  deviations $\delta\left(e^{2\V} \right)$, $\delta\left(\W \right)$ and the parameter $a^2$ are small, 
the equations can be linearized. 
This gives inhomogeneous  
equations for $\delta\left(e^{2\V} \right)$, $\delta\left(\W \right)$ 
with 
source term proportional to $a^2$. Consequently, the deviations are expected  to scale as  $a^2$. 
\\
\indent
We checked that the linearized  equations admit a bounded solution satisfying, at 
 the throat, 
\be
\left.\delta\left(e^{2\V} \right)\right|_{x=0}=a^2\times \frac{1+y^2}{4}\,,~~~~~~~
\frac{\partial }{\partial x}\left. \delta\left(e^{2\V}\right) \right|_{x=0}=0\, ,~~~~~~~
\left. \delta\left(\W \right) \right|_{x=0}=\mathrm{const.}\, 
\ee
The second and third of these relations follow from the wormhole boundary conditions, while the first one 
can be obtained  by a local analysis in the vicinity of $x=0$. 
The profiles of  $\delta\left(e^{2\V} \right)/a^2$ and  $\delta\left(\W \right)/a^2$ 
are similar to those shown in Fig.\ref{Fig66}. 
\\
\indent
To summarize, the above considerations indicate that the 
 rapidly spinning wormholes converge  to the extremal Kerr configuration as $a\to 0$. 
 However, 
one should emphasize that this statement applies only to the  $x>0$ region. 
}

Let us discuss the extension to the negative values of $x$.
The wormholes are symmetric under $x\to -x$, up to a flip in the sign of the rotation field,
\be
\V(-x,y)=\V(x,y),~~~~\W(-x,y)=-\W(x,y).
\ee
The Kerr solution \eqref{Kerr3} remains invariant, up to the same sign flip in the rotation, if the reflection $x\to -x$
is accompanied by a reflection of the mass parameter $M\to -M$,
\be
\V_{\rm K}(-x,y,-M)=\V_{\rm K}(x,y,M),~~~~\W_{\rm K}(-x,y,-M)=-\W(x,y,M).
\ee
This is an isometry, and the ADM mass seen from $x\to \pm\infty$ is the same.
Therefore, the spinning wormhole configurations $\V(x,y)$ and $\W(x,y)$ 
approach $\V_{\rm K}(x,y,M)$ and $\W_{\rm K}(x,y,M)$ for $x>0$, and
$\V_{\rm K}(-x,y,-M)$ and $\W_{\rm K}(-x,y,-M)$ for $x<0$.
In other words, the wormholes approach the extremal Kerr geometry for $x>0$ and its isometric counterpart for $x<0$. 
In both cases, only the exterior part of the black hole geometry is involved.

\blue{
One should emphasize that the limiting configuration consisting of two black-hole exteriors, 
although it approximates the wormholes, is not itself a global solution of the field equations. 
Indeed, the exterior region of the Kerr geometry analytically continues for $x<0$ into the interior, 
rather than into another exterior region. 
For example, the trajectory $x(\tau)$ of a particle falling along the symmetry axis, where $y=1$, 
is described by 
\be
\left( \frac{dx}{d\tau} \right)^2+e^{2\V(x,1)}={\cal E}^2, 
\ee
where $V$ corresponds either to the wormhole or to the Kerr geometry. 
In the latter case, this equation is precisely the same as \eqref{radg}. 
In both cases, one has $dx/d\tau<0$ at $x=0$, so the trajectory coming from the $x>0$ region 
continues into the $x<0$ region. 
\\
\indent 
Consider the Schwarzschild  radius $r_{\rm S}=\rho\, e^{-V}$  defined by 
\eqref{Schw}, whose derivative is 
$dr_{\rm S}/d\tau=(dr_{\rm S}/dx)(dx/d\tau)$. 
For the extremal black hole, one has on the axis $r_{\rm S}=x\, e^{-\V_{\rm K}(x,1)}$ and $\left.dr_{\rm S}/dx\right|_{x=0}=1/\sqrt{2}$, hence 
$\left.dr_{\rm S}/d\tau\right|_{x=0}<0$. 
Therefore, 
in the Kerr geometry, $r_{\rm S}$ 
continues to decrease as the geodesic crosses the horizon at $x=0$, so that the 
geodesic proceeds into the interior, where $r_{\rm S}<\left.r_{\rm S}\right|_{x=0}$, rather than into another exterior region. 
Thus, joining two Kerr exteriors at $x=0$ produces a geodesically incomplete manifold, 
since geodesics leave the manifold through the joining surface.
\\
\indent 
The wormholes approaching this limiting configuration, however, are geodesically complete. 
In this case, $r_{\rm S}=\sqrt{x^2+a^2}\, e^{-V(x,1)}$, and  
$\left.dr_{\rm S}/dx\right|_{x=0}=\left.dr_{\rm S}/d\tau\right|_{x=0}=0$. Hence, $r_{\rm S}$ attains a minimum at $x=0$ 
and then starts to increase. 
A geodesic arriving from one asymptotic region therefore continues through the 
throat into another asymptotic region, whose geometry is also approximately Kerr.
\\
\indent 
In summary, the wormholes are global solutions of the field equations and are geodesically complete, since their geodesics can be extended indefinitely. In the limit $a\to0$, they globally approach a configuration obtained by joining two exterior regions of the extremal Kerr geometry. This limiting configuration, however, solves the field equations only locally and is not geodesically complete. The limiting configuration can therefore be approached but never attained, reflecting the fact that the speed of light, $v_0\to1$, can be approached but not reached.
}

Since the rapidly spinning wormhole is seen from both sides as an almost extremal Kerr geometry, 
it effectively ``mimics'' a black hole \cite{Damour:2007ap}. In other words, the extremal Kerr black-hole 
geometry is imitated by a wormhole geometry produced by a spinning ring whose size in the static limit 
tends to zero, while its rotational velocity approaches unity, 
stretching the ring to a finite size equal to the Kerr equatorial radius.

As long as the limit is not reached, the wormholes contain a singularity at the ring, 
but it disappears in the limit since the ring tension parameter $\eta$ defined by \eqref{Kr2o} 
approaches zero, as seen in Fig.\ref{Fig3}.
For the extremal Kerr solution \eqref{Kerr3} itself one has
\be
\left.\partial_x W_K\right|_{x=0}=-\frac{1}{2 M^2}\,,
\ee
hence, using \eqref{Rn}, one obtains
\be                \label{Kr2}
\C=1-\left.\frac{R_0^4}{4}\, \W_{,x}^2\right|_{x=y=0}=1-\frac{R_0^4}{4}\left( \frac{1}{2 M^2}\right)^2
=1-\frac{(2 M)^4}{4}\left( \frac{1}{2M^2}\right)^2=0.
\ee
According to \eqref{Cp}, one has $C_\phi=\sqrt{\eta}=0$, 
hence the scalar field amplitude for the dressed solutions vanishes as well.
As a result, the fast rotation limit is the same for the vacuum and dressed 
wormholes, so that  the latter also approach the Kerr 
geometry in the limit 
\cite{Kleihaus:2014dla,Chew:2016epf}.

 \section{Properties of wormholes \label{SctIX}}

 The wormhole throat is characterized by the equatorial radius determined by the length of its equatorial circumference,
\be
R_0&=&\frac{1}{2\pi} \int_0^{2\pi} \left.\sqrt{g_{\varphi\varphi}}\right|_{x=y=0}\,d\varphi=a\, e^{-V_0}.
\ee

Since the throat is not spherically symmetric, one can also introduce a polar radius determined by the length of the polar circumference,
\be             \label{Rp}
R_p&=&\frac{2}{\pi} \int_0^1 \left.\sqrt{g_{yy}}\right|_{x=0}\,dy
=\frac{2a}{\pi} \int_0^1 \left.e^{\K-V}\right|_{x=0}\,\frac{dy}{\sqrt{1-y^2}}\,.
\ee

In addition, one can define the average radius determined by the throat area,
\be                \label{RA}
R_A^2&\equiv& \frac{A}{4\pi}=\frac{1}{2\pi}\int_0^{2\pi} d\varphi\int_0^1 \left.\sqrt{g_{yy}g_{\varphi\varphi}}\right|_{x=0}\, dy=
\nuu  \int_0^1 \left.e^{\K-2V}\right|_{x=0}\,dy\,.
\ee

Unlike $R_0$, the radii $R_p$ and $R_A$ involve the metric amplitude $\K$, 
and therefore depend on whether the wormhole is vacuum or scalar-dressed.

As the wormhole spins faster, the ratios $R_p/R_0$ and $R_A/R_0$ decrease, and the configuration becomes more oblate.
 
 \begin{figure}[th]
\hbox to \linewidth{ \hss

			\includegraphics[width=8 cm]{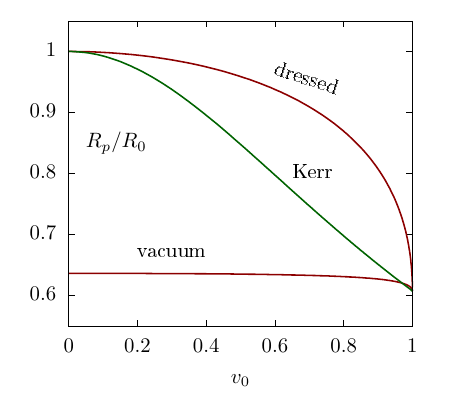}	
	\includegraphics[width=8 cm]{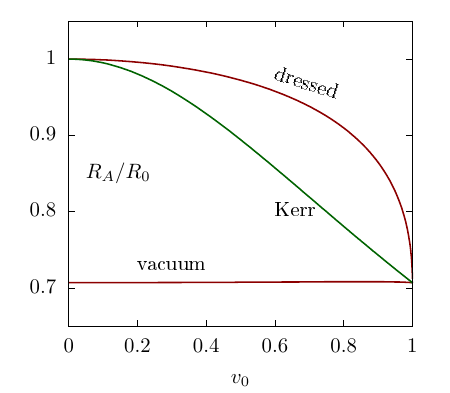}

\hss}

\caption{The ratios $R_p/R_0$ (left) and $R_A/R_0$ (right) for the vacuum and scalarized  wormholes
as functions of the throat linear velocity, and for the Kerr black hole as a function of the horizon linear velocity.}
\label{Fig7}
\end{figure}

 The numerical curves for $R_p/R_0$ and $R_A/R_0$ for vacuum and scalar-dressed wormholes
as functions of the throat linear velocity $v_0$ are shown in Fig.\ref{Fig7}, together with the
corresponding curves for the Kerr black hole defined by Eqs.\eqref{Rpa},\eqref{RAa} in \ref{AppB}.
The dressed wormholes become spherically symmetric in the static limit, with $R_p/R_0\to 1$ as $v_0\to 0$.

The throat of the vacuum wormhole reduces in the static limit to a disk, whose circumference is 
$2\pi R_0=2\pi a$, while the “polar circumference” is twice the diameter, $4R_0=4a$. 
Their ratio $4R_0/(2\pi R_0)=2/\pi\approx 0.636$ corresponds to the value of $R_p/R_0$ in the limit $v_0\to 0$.

For rapid rotation, both vacuum and dressed wormholes approach the extremal Kerr solution, 
and hence $R_p/R_0\to 0.608$, in agreement with Eq.\eqref{Rpa}.
  
 The throat area of the vacuum wormhole  in the static limit is twice the disk area, 
 $A=2\times \pi R_0^2\equiv 4\pi R_A^2$. 
 Therefore,  $R_A/R_0\to 1/\sqrt{2}\approx 0.707$ for $v_0\to 0$. 
 For the rapid rotation  both the vacuum and dressed wormholes 
 approach the extreme Kerr metric, hence  $R_A/R_0\to 1/\sqrt{2}$, as explained by  \eqref{RAa}. 
 
 For the vacuum wormholes the ratio $R_A/R_0$ is almost constant 
 and approaches  $1/\sqrt{2}$ for $v_0\to 0,1$, but 
 zooming the curve reveals  that $R_A/R_0$ is not exactly constant.

 \subsection{Embedding diagrams}

 The geometry of the wormhole throat is similar to that of a deformed 2-sphere.
It is described by the line element obtained by setting in \eqref{METR} $t=\mathrm{const.}$ and $x=0$, hence
\be            \label{7.19}
ds^2=\frac{\nuu \, e^{2K-2V} }{1-y^2}\, dy^2+\nuu \, e^{-2V} (1-y^2) d\varphi^2
\equiv g_{yy}(y) dy^2+g_{\varphi\varphi}(y) d\varphi^2,
\ee
where $K,V$ are evaluated at $x=0$.

One can try to realize the same geometry as a surface of revolution in $\mathbb{R}^3$ determined 
by the Euclidean coordinates
$X={\cal R}(y)\cos\varphi$, $Y={\cal R}(y)\sin\varphi$, and $Z={\cal Z}(y)$, hence
\be              \label{ds0}
ds^2=dX^2+dY^2+dZ^2=\left( {\cal R}^{\prime 2}+Z^{\prime 2}  \right)dy^2+{\cal R}^2(y) d\varphi^2.
\ee 
 Comparing with \eqref{7.19} yields
\be          \label{lb}
{\cal R}(y)=\sqrt{g_{\varphi\varphi}},~~~~~
Z^\prime(y)=\sqrt{g_{yy}- {\cal R}^{\prime 2} }.
\ee
Applying these formulas in the static limit described by \eqref{ring},\eqref{ring1} yields
${\cal R}(y)=\sqrt{1-y^2}$, while $Z^\prime=0$ for the vacuum solution
and $Z^{\prime 2}=1$ for the dressed solution.
The line element \eqref{ds0} reduces to
\be
\text{vacuum:}~~~ds^2&=&d{\cal R}^2+{\cal R}^2d\varphi^2,~~~~~
\text{dressed:}~~~ds^2=d\vartheta^2+\sin^2\vartheta d\varphi^2\,,
\ee
where in the latter case we have set $y=\cos\vartheta$. In the vacuum case this
corresponds to a two-sided disk, because ${\cal R}(y)=\sqrt{1-y^2}$ is covered twice for $y\in[-1,1]$,
while for the dressed solution this describes a 2-sphere.
 
 \begin{figure}[th]
\hbox to \linewidth{ \hss	
			\includegraphics[width=8.5 cm]{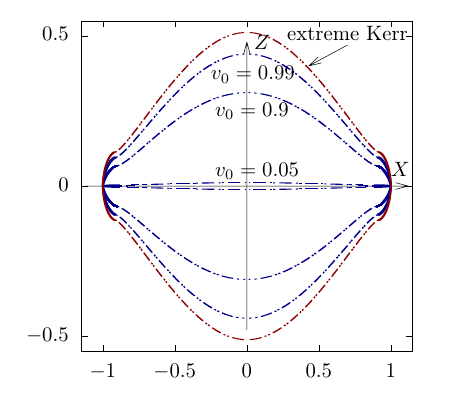}	
	\includegraphics[width=8.5 cm]{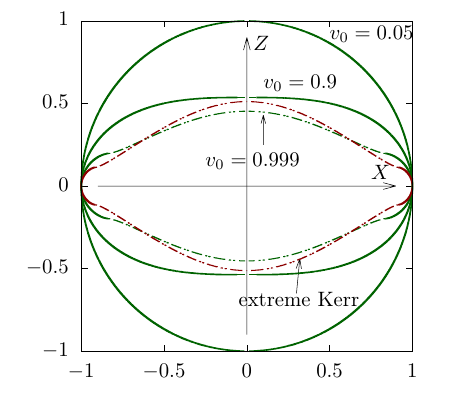}	
\hss}
\caption{The isometric embeddings of the wormhole throat for the vacuum (left) and scalar-dressed (right) solutions
correspond to surfaces in the $X,Y,Z$ space obtained by rotating the curves around the $Z$-axis.
The solid parts of the curves denote regions that can be embedded
into $\mathbb{R}^3$,
while the dashed parts correspond to regions embedded into $\mathbb{R}^{1,2}$. All solutions correspond to $R_0=1$.
}
\label{Fig8}
\end{figure}

For the spinning vacuum wormholes the throat geometry describes a ``thick'' disk, but only its
edge, where $y$ is close to zero,
can be embedded into $\mathbb{R}^3$. For larger values of $y$ the radicand in \eqref{lb} becomes negative, and
the corresponding part of the 2-surface
cannot be isometrically embedded into $\mathbb{R}^3$, but can instead be embedded into the pseudo-Euclidean space
$\mathbb{R}^{(1,2)}$.

A similar phenomenon is known for the horizon geometry of rapidly spinning Kerr black holes, which can be partly
embedded into $\mathbb{R}^3$ and partly into $\mathbb{R}^{(1,2)}$ \cite{Smarr:1973zz} (see \ref{AppA}).

 Therefore, one generalizes \eqref{ds0},\eqref{lb} to allow for a different metric signature of the ambient space,
\be              \label{ds}
ds^2&=&dX^2+dY^2\pm dZ^2=\left( {\cal R}^{\prime 2}\pm Z^{\prime 2}  \right)dy^2+{\cal R}^2(y) d\varphi^2 \nn \\
{\cal R}(y)&=&\sqrt{g_{\varphi\varphi}},~~~~~
Z^\prime(y)=\sqrt{\pm(g_{yy}- {\cal R}^{\prime 2})}.
\ee
One chooses ``+'' where $g_{yy}> {\cal R}^{\prime 2}$ and ``-'' otherwise, which makes
the radicand always non-negative.

The resulting profiles $X(y)=\pm{\cal R}(y)$ and $Z(y)$ of the surfaces of revolution are
shown in Fig.\ref{Fig8} (left panel). The solid parts of the curves correspond to the ``+'' regions,
which can be embedded into
$\mathbb{R}^3$, while the dashed parts denote the ``-'' regions
embedded into $\mathbb{R}^{1,2}$.
As a result, the total geometry can be viewed as a combination of a sphere and a hypersphere.

As seen in Fig.\ref{Fig8}, only a small equatorial part of the throat can be embedded into $\mathbb{R}^3$.
Most of the throat geometry can be embedded only into $\mathbb{R}^{1,2}$, 
giving the erroneous impression that rapidly rotating
configurations become thick and that the polar regions bulge. 
However, as seen in Fig.\ref{Fig7}, for vacuum solutions the oblateness
determined by the ratios $R_p/R_0$ and $R_A/R_0$ is almost independent of the rotation and remains
essentially the same as for the thin disk.

Notice also that in the fast rotation limit the embedding diagram for the throat geometry approaches that for the horizon
geometry of the extremal Kerr black hole (see Fig.\ref{Fig10} in \ref{AppA}).

The scalar-dressed solutions shown in the right part of Fig.\ref{Fig8} exhibit properties that are, in a sense, opposite to
those of the vacuum solutions. In the static limit the throat is a perfect sphere, which becomes squashed
under slow rotation, although the curvature remains positive and the embedding into $\mathbb{R}^3$ is still possible.

For $v_0\approx 0.9$ the curvature vanishes at the poles,
and for faster rotation the polar regions can be embedded only into $\mathbb{R}^{1,2}$. In the $v_0\to 1$ limit the
embedding diagram approaches that of the extremal Kerr geometry. As seen in Fig.\ref{Fig7}, the oblateness always grows
with the rotation.

 \begin{figure}[th]
\hbox to \linewidth{ \hss	
			\includegraphics[width=8.5 cm]{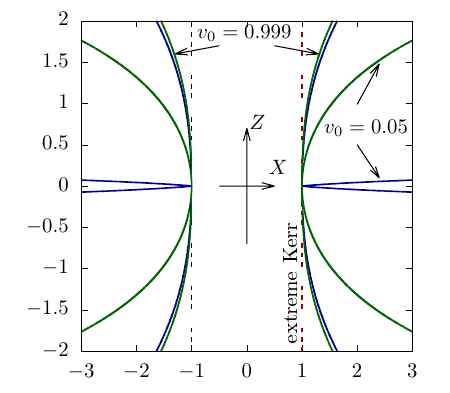}	

\hss}
\caption{
The isometric embeddings of the equatorial sections of wormholes with $R_0=1$
correspond to surfaces in the $X,Y,Z$ space obtained by rotating the curves around the $Z$-axis, which are either 
blue or green for the vacuum and dressed solutions, respectively.
For slow rotation, the surface for the vacuum solution is close to the $Z=0$ plane
with a unit disk removed, while that
for the dressed solution is close to the catenoid $\sqrt{X^2+Y^2}=\cosh Z$.
For rapid rotation all surfaces
approach the cylinder $X^2+Y^2=1$ corresponding to the near-horizon limit of the extremal Kerr black hole.
}
\label{Fig9}
\end{figure}

 Let us finally consider the geometry of the equatorial section.
Setting in \eqref{METR} $t=\mathrm{const.}$ and $y=0$ yields
\be
ds^2=e^{2K-2V}dx^2+e^{-2V}(x^2+\nuu ) d\varphi^2\equiv g_{xx}(x) dx^2+g_{\varphi\varphi}(x) d\varphi^2\,,
\ee
where $K,V$ are evaluated at $y=0$. This geometry can be isometrically embedded
into $\mathbb{R}^3$ by the same procedure as above,
\be              \label{dso}
ds^2=dX^2+dY^2+dZ^2=\left( {\cal R}^{\prime 2}+Z^{\prime 2}  \right)dx^2+{\cal R}^2(x) d\varphi^2,
\ee
with ${\cal R}(x)=\sqrt{g_{\varphi\varphi}}$ and
$Z^\prime(x)=\sqrt{g_{xx}- {\cal R}^{\prime 2} }$. This time the radicand is always non-negative.
 
 In the static limit this gives
\be
\text{vacuum:}~~~ds^2&=&d{\cal R}^2+{\cal R}^2d\varphi^2,~~~~~~~~~~~~~~~{\cal R}=\sqrt{x^2+1}\geq 1,~~~ Z=0;\nn \\
\text{dressed:}~~~
ds^2&=&\cosh^2 Z\,(dZ^2+d\varphi^2)\,,~~~~~{\cal R}=\cosh Z,~~x=\sinh Z.
\ee
In the vacuum case
this corresponds to a doubly covered $X,Y$ plane with a unit disk removed.
In the dressed case this is a catenoid.

As seen in Fig.\ref{Fig9}, for the
slowly rotating solutions the doubly covered plane
splits into two sheets that start opening up.
For rapid rotation the catenoid also opens up, the surfaces for the vacuum and dressed solutions approach each other,
and both approach the equatorial section of the Kerr geometry, which converges in the extremal limit to a cylinder 
 in the $X,Y,Z$ space.

\section{Summary and concluding remarks \label{SctX}}
\setcounter{equation}{0}

We constructed above the spinning generalizations of the static ring wormholes obtained from the Kerr black hole
in the zero-mass limit. The static wormhole is locally flat and represents two copies of Minkowski space 
glued together through a 2-disk
whose radius $a$ can be arbitrary. The disk is encircled by a ring carrying a conical singularity
with a negative angle deficit $\Delta\psi=-2\pi$, which can be viewed as a loop made of a cosmic 
string with negative tension $T=-1/4$.
Therefore, even though the wormhole is described by a vacuum metric, there is negative energy
hidden in the ring singularity.

The spinning wormholes are no longer locally flat. Their geometry is not Kerr,
and they are
invariant under reflections across the wormhole throat, up to a flip in the sign of the angular momentum.
Their ADM mass is positive and is the same when viewed from both infinities.

The spinning solutions can be labeled by the size parameter $a\in (0,\infty)$ determining
the ring radius in the static limit,
and by the angular momentum $J\in[0,\infty)$.

A slowly rotating wormhole satisfies the non-relativistic relation between $M$ and $J$,
\be
M=\frac{J^2}{2I},~~~~~~~J\ll 1,
\ee
which transforms into the Regge relation in the fast-rotation regime,
\be
J=M^2,~~~~~~J\gg 1.
\ee
The spinning ring is stretched by the centrifugal force, and its radius $R_0$ increases up to a value
depending on the angular momentum,
\be
a\leftarrow R_0 \to 2\sqrt{J}~~~~~\text{as}~~~~0\leftarrow J\to \infty.
\ee
Instead of the angular momentum $J$,
the wormhole rotation can be characterized by the linear velocity of the ring, $v_0\in [0,1)$. 
When $v_0$ increases, $J$ grows,
whereas the ring tension parameter $\eta$ decreases 
from its unit value to zero and determines the tension,
\be
T=-\frac{\C}{4}.
\ee
Interestingly, the product $\eta R_0$ changes very little with the rotation,  
$a\leftarrow\eta R_0\to 4a/\pi$, hence one can define the ring energy
\be
E_{\rm ring}=2\pi R_0\times T=-\frac{\pi}{2}\,\eta R_0\approx -\frac{\pi}{2}\,a,
\ee
which is approximately the same for the static and spinning rings.

The fast-rotation limit corresponds to $v_0\to 1$, in which case $M$, $J$, and $R_0$
grow without bounds, provided that $a$ is fixed. 
This corresponds to a ring that spins faster and faster, stretching in size and
increasing its energy and angular momentum more and more. However, one can take the $v_0\to 1$ 
limit while simultaneously sending
the size parameter $a$ to zero, in which case $M$, $J$, and $R_0$ remain finite.

For example, one can require the ring radius to be independent of the rotation,
$R_0=2\m=\mathrm{const.}$ This determines a family of rings labeled by their size in the static limit, $a\leq 2\m$.
Each ring spins with
the linear velocity $v_0(a)$ adjusted such that, for any $a$, it stretches to the same radius $R_0=2\m$.

The mass and angular momentum for such solutions vary within finite limits,
$M\in [0,\m)$, $J\in [0,\m^2)$, and in the fast-rotation limit one has $M\to \m$, $J\to M^2$, 
which again reproduces the Regge relation.

The wormhole geometry on each side of the spinning ring with fixed $R_0=2{\cal M}$ 
approaches for $v_0\to 1$  the exterior region
of the extremal Kerr geometry
with mass ${\cal M}$. Therefore, the wormholes mimic the Kerr geometry.

The vacuum solutions carry a curvature singularity consisting of a distributional part located at the ring and a volume
part associated with components of the Riemann tensor that diverge when approaching the ring. The singularity
plays the role of an effective source of negative energy needed for the wormhole to exist.
However, the singularity can be removed via the scalarization procedure by adding a phantom scalar field.
In this case the solutions describe spinning generalizations of the ultra-static
BE wormhole \eqref{BE0}.

Such solutions were studied in Refs.\cite{Kleihaus:2014dla,Chew:2016epf},
and we were able to verify all
results of that analysis within  our approach, finding only a couple of minor corrections.
First, it was stated in \cite{Kleihaus:2014dla,Chew:2016epf}
that the angular momentum is symmetric under $x\to -x$, whereas we know that it is antisymmetric.
Second, we obtain the same curve for the angular momentum 
as the one shown in Fig.2 of \cite{Chew:2016epf}
if we plot $2\times J/R_0^2$ rather than $J/R_0^2$, as stated in that paper. In any case,
$J/R_0^2$ approaches $1/4$ and not $1/2$ in the fast-rotation limit.

Before concluding, one should mention that the original motivation for 
this work was the search for spinning solutions in a closed analytic form.
One may expect that the spinning
generalization of the static ring wormhole should be obtainable analytically. 
However, applying various solution-generating techniques did not lead to success
\cite{Volkov:2021blw}. Analytical solutions can be obtained, 
but they are not satisfactory for various reasons:
either they are not asymptotically flat, 
or they contain additional line singularities of the Taub-NUT type,
or they exhibit other problems. Asymptotically flat spinning wormholes 
containing  only the ring singularity are also known,
but they are not vacuum solutions \cite{Clement:1998nk}.

This is why we used numerical methods to clarify the structure of the solutions, which may be interesting
in their own right. In addition, this analysis may be helpful for future attempts to construct the solutions analytically,
perhaps by applying the inverse-scattering method \cite{Belinski:2001ph} 
or by using some other approach \cite{Hoenselaers:1979mk}.

{\bf Acknowledgements.} It is a pleasure to thank Sergey Solodukhin for discussions and Romain Gervalle for help 
with FreeFem++ during the early stages of this work.

\renewcommand{\thesection}{Appendix A}
\section{Dual forms of the metric and the  Ernst equations\label{AppA}}

\renewcommand{\theequation}{A.\arabic{equation}}
\setcounter{equation}{0}
\setcounter{subsection}{0}

We summarize in this Appendix and in the two Appendices that follow
some technical results obtained using the Ernst formalism that are employed in the main text.

To begin with, we describe the two choices of metric variables associated either with the timelike
or with the axial Killing vector. One of them corresponds to the variables 
used to describe wormholes, while the other
is more appropriate for describing the Kerr metric.

The line element \eqref{METR}  in the main text can be represented in the two equivalent forms, 
\begin{subequations} 
\begin{align}               \label{METRxx}
ds^2
&=-e^{2\V}dt^2+e^{-2\V} \left\{e^{2 {\K}}\, dh^2 +\rho^2 (d\varphi-\W\, dt)^2\right\} , \\
&=-e^{2U}(dt-w\,d\varphi)^2+e^{-2U}\left\{e^{2{k}} dh^2+ \rho^2 d\varphi^2 \right\}\,,  \label{METRxxx}
\end{align} 
\end{subequations} 
with 
\be
dh^2=dx^2+\frac{x^2+\nu }{1-y^2}\, dy^2\,,~~~~~\rho^2=(x^2+\nu )(1-y^2). 
\ee
For wormholes one has
$
\nu=a^2,
$
but we now use the symbol $\nu$ in order to include black holes as well, for which $\nu$ is non-positive.

The form \eqref{METRxx} is expressed in terms of the amplitudes $\V,\W,\K$, 
while \eqref{METRxxx} is parameterized by $U,w,k$,
and the metric components are
\be              \label{UV}
g_{00}&=&-e^{2U}=-e^{2\V}+\rho^2\W^2 e^{-2\V}\,,~~~\nn \\
g_{0\varphi}&=&we^{2U}=-\rho^2\W e^{-2\V},~~~\nn \\
g_{\varphi\varphi}&=&\rho^2 e^{-2U}-w^2 e^{2U}=\rho^2 e^{-2\V}\equiv e^{2{\rm U}}\,, \nn \\
g_{xx}&=&e^{2k-2U}=e^{2\K-2\V}\,, \nn \\
g_{yy}&=&\frac{x^2+\nu }{1-y^2}\times g_{xx}.
\ee
We shall call \eqref{METRxx} and \eqref{METRxxx} dual, or conjugate, forms of the same line element.

The form \eqref{METRxx}, expressed in terms of $\V$, will be called the
$\varphi$-parametrization,
because $e^{2\rm U}=\rho^2 e^{-2\V}$ is the norm of the axial Killing vector $\partial_\varphi$.
This parametrization is used in the main text to describe wormholes. On the other hand, the Kerr solution
is more conveniently described by the line element \eqref{METRxxx} expressed in terms of $U$.
One can call this the $t$-parametrization,
because $e^{2U}$ is the norm of the timelike Killing vector $\partial_t$.

As a result, one can choose as the basic variables either the functions ${\rm U}=\ln\rho-V,\W$ or the functions $U,w$.
They are nontrivially related to each other via \eqref{UV},
\be               \label{WwUu}
e^{2{\rm U}}=\rho^2 e^{-2U}-w^2 e^{2U},~~~~~\W e^{2{\rm U}}=-w e^{2U}.
\ee
However, ${\rm U},\W$ and $U,w$ satisfy exactly the same equations.
Specifically, setting $\V=\ln\rho-{\rm U}$ in Eq.\eqref{UUU} in the main text yields the ${\rm U},\W$ equations,
\be
\label{UU}
[(x^2+\nu ){\rm U}_{,x}]_{,x}+[(1-y^2){\rm U}_{,y}]_{,y}+\frac{1}{2\rho^2}\,e^{4{\rm U}}\left[
(x^2+\nu ) \W_{,x}^2+(1-y^2)\W_{,y}^2
\right]&=&0, \nn \\
\left[\frac{e^{4{\rm U}}}{\rho^2}(x^2+\nu )\W_{,x}\right]_{,x}+\left[\frac{e^{4{\rm U}}}{\rho^2}(1-y^2)\W_{,y}\right]_{,y}&=&0,
\ee
while replacing here ${\rm U}\to U$ and $\W\to w$ yields the equations for $U,w$.

Similarly, setting $\V=\ln\rho-{\rm U}$ in Eqs.\,\eqref{KK0} for $\K$ in the main text and then replacing
${\rm U}\to U$ and $\W\to w$ yields the equations for $k$,
\be               \label{KK00}
\partial_x k&=&\frac{1-y^2}{x^2+\nu\,y^2}\left( \Gamma(U)
-\frac{e^{4U}}{4\rho^2 }\,\Gamma(w)+\frac{\nu\, x}{x^2+\nu}
\right), \nn \\
\partial_y k&=&\frac{x^2+\nu}{x^2+\nu\,y^2}\left( \Lambda(U)
-\frac{e^{4U}}{4\rho^2 }\,     \Lambda(w)+\frac{\nu\, y}{x^2+\nu}
\right).
\ee

The same correspondence works if we replace the rotation fields $\W$ and $w$ 
by the corresponding twist potentials.
The twist potential $\chioo$ for $\W$ is defined by Eqs.\,\eqref{twisto} in the main text. 
Re-expressing these relations in terms of ${\rm U}$, one obtains
\be            \label{ww}
\W_{,x}=(y^2-1)e^{-4{\rm U}}\chioo_{,y},~~~~\W_{,y}=(x^2+\nu )e^{-4{\rm U}}\chioo_{,x},
\ee
and inserting these expressions into \eqref{UU} yields the ${\rm U},\chioo$ equations,
\be
\label{Ua}
[(x^2+\nu ){\rm U}_{,x}]_{,x}+[(1-y^2){\rm U}_{,y}]_{,y}+\frac12\,e^{-4{\rm U}}\left[
(x^2+\nu ) \chioo_{,x}^2+(1-y^2)\chioo_{,y}^2
\right]&=&0, \nn \\
\left[e^{-4{\rm U}}(x^2+\nu )\chioo_{,x}\right]_{,x}+\left[e^{-4{\rm U}}(1-y^2)\chioo_{,y}\right]_{,y}&=&0.
\ee
The twist potential $\chio$ for the $w$ field is defined by exactly the same relations as in \eqref{ww},
\be            \label{www}
w_{,x}=(y^2-1)e^{-4U}\chio_{,y},~~~~w_{,y}=(x^2+\nu )e^{-4U}\chio_{,x},
\ee
and the equations for $U,\chio$ are obtained by replacing ${\rm U}\to U$ and $\chioo\to\chio$ in \eqref{Ua}.

Introducing the complex Ernst potentials \cite{Ernst}, constructed either from ${\rm U},\chioo$ or from $U,\chio$,
\be            \label{Er1}
{\cal E}_{(\varphi)}=e^{2{\rm U}}+i\chioo\equiv  \frac{\xi-1}{\xi+1},~~~~~~~~~
{\cal E}_{(t)}=e^{2U}+i\chio\equiv  \frac{\xi-1}{\xi+1},
\ee
the equation for the complex function $\xi$ is the same in both cases: 
\be                \label{Er}
(\xi\bar{\xi}-1)\left\{~[(x^2+\nu )\xi_{,x}]_{,x}+[(1-y^2)\xi_{,y}]_{,y}~\right\}=2\bar{\xi}\,\left[
(x^2+\nu )\xi_{,x}^2+(1-y^2)\xi_{,y}^2
\right].
\ee

\renewcommand{\thesection}{Appendix B}
\section{Kerr solution \label{AppB}}

\renewcommand{\theequation}{B.\arabic{equation}}
\setcounter{equation}{0}
\setcounter{subsection}{0}

We describe here the Kerr solution and some of its features used in the main text. To begin with,
the complex Ernst equation \eqref{Er} admits a simple solution \cite{Ernst},
\be                  \label{Kerrxi}
\xi=px-iqy~~~~~~\text{with} ~~~~~~~~\nu=\frac{q^2-1}{p^2},
\ee
where $p,q$ are real parameters.
Inserting this into \eqref{Er1} determines the Ernst potential in the $t$-parametrization, ${\cal E}_{(t)}=e^{2U}+i\chio$,
from which one obtains $U$ and $\chio$.
Using \eqref{KK00} and \eqref{www}
to compute $k$ and $\chio$ yields the amplitudes in the $t$-parametrization, 
\bea                \label{xr}
e^{2U}&=&1-\frac{2(px+1)}{(px+1)^2+q^2 y^2},~~~~~~~~\chio=-\frac{2qy}{(px+1)^2+q^2 y^2}, \nn \\
w&=&\frac{2q}{p}\times \frac{(px+1)(y^2-1)}{p^2 x^2+q^2 y^2-1},~~~~~
e^{2k}=\frac{p^2 x^2+q^2 y^2-1 }{p^2\,(x^2+\nu )}.
\eea
\blue{
Substituting these into \eqref{UV} gives the amplitudes in the $\varphi$-parametrization,
 \be              \label{prep}
e^{2\V}&=&\frac{[\,p^2 x^2+q^2-1][\,(px+1)^2+q^2 y^2] }{\cal Q},~~~~~\W=\frac{2\,q\,p\,(px+1)}{\cal Q},  \nn \\
e^{2\K}&=&\frac{[(px+1)^2+q^2 y^2]^2 }{\cal Q},~~~~~~~
\omega=\frac{2\,q\,y\,[(px+1)^2(3-y^2)+q^2(1+y^2) ] }{p^2\, [(px+1)^2+q^2 y^2 ]}, \nn \\
{\cal Q}&=&(px+1)^4+q^2\,(px+1)[\,px+3+(px-1)\,y^2 ]+q^4y^2. 
\ee
These  expressions determine the complex potential 
${\cal E}_{(\varphi)}=\rho^2 e^{-2V}+i\omega=(\xi-1)/(\xi+1)$,  
where $\xi$  also satisfies   the Ernst equation \eqref{Er},  although it is quite different from the one in \eqref{Kerrxi}. 
 We use these $\varphi$-amplitudes  to compare with the wormhole solutions, in particular in the extremal limit $q=1$. 
Notice that the twist $\omega$ satisfies the 
boundary conditions \eqref{bc} both at the equator and on the axis: $\omega(x,0)=0$ and  $\omega(x,1)=4q/p^2=4aM=4J$. 
}

Inserting \eqref{xr} or \eqref{prep}  into \eqref{UV} and defining
\be                \label{B3}
p=\frac{1}{M},~~~~r=x+M,~~~~q=\frac{a}{M},~~~~\rK=r^2+a^2 y^2\,,~~~~\Delta=x^2+\nu,
\ee
yields the metric components
\be                  \label{Kcomp}
g_{00}&=&-1+\frac{2Mr}{\rK},~~~g_{xx}=\frac{\rK}{\Delta},~~~~g_{yy}=\frac{\rK}{1-y^2}, \nn \\
g_{0\varphi}&=&-\frac{2aMr}{\rK}\,(1-y^2),~~~~~g_{\varphi\varphi}=\left(r^2+a^2-a\times g_{0\varphi} \right)(1-y^2).
\ee

The line element
$
ds^2=g_{\mu\nu} dx^\mu dx^\nu
$
assumes the form
\be                  \label{Kmtr}
ds^2=-dt^2+\frac{\rK\,dx^2}{\Delta}\, +\frac{\rK\, dy^2}{1-y^2}\,+(r^2+a^2)(1-y^2)\, d\varphi^2
+\frac{2Mr}{\Sigma}\left(dt -a\,(1-y^2)\, d\varphi\right)^2.~~~~~
\ee
Setting finally
\be                        \label{rr}
M=\frac{\rm M}{\MU},~~~
t=\frac{\rm t}{\MU},~~~
r=\frac{\rm r}{\MU},~~~
x=r-M,~~~
y=\cos\vartheta,~~~~
\ee
the dimensionful metric $d{\rm s}^2=\MU^2 ds^2$ reduces precisely to the Kerr line element \eqref{Kerr-m}.

It is also worth noting that, introducing two 1-forms and two vectors,
\be
\theta_1 &=&dt-a\,(1-y^2)\, d\varphi,~~~~\theta_2=(r^2+a^2)\,d\varphi-a\,dt,\nn \\
\eta_1&=&(r^2+a^2) \partial_t+a\,\partial_\varphi,~~~~
\eta_2=\partial_\varphi+a\,(1-y^2)\,\partial_t,
\ee
such that $\langle \theta_a,\eta_b \rangle=\Sigma\, \delta_{ab}$, one has
\be
g_{\mu\nu} dx^\mu dx^\nu&=&-\frac{\Delta}{\Sigma}\,\theta_1^2+\frac{1-y^2}{\Sigma}\,\theta_2^2
+\frac{\Sigma}{\Delta}\, dx^2+\frac{\Sigma}{1-y^2}\, dy^2\,, \label{Kmtr1} \\
g^{\mu\nu} \partial_\mu \partial_\nu&=&-\frac{1}{\Delta \Sigma}\, \eta_1^2+\frac{1}{\Sigma\,(1-y^2)}\,\eta_2^2
+\frac{\Delta}{\Sigma} (\partial_x)^2+\frac{1-y^2}{\Sigma} (\partial_y)^2\,.
\ee

\subsubsection{Event horizon}

The black hole horizon is located where $x^2+\nu =0$ and $g_{xx}\to\infty$,
hence at
\be
x_{h}=\sqrt{-\nu }=\frac{\sqrt{1-q^2}}{p}=\sqrt{M^2-a^2}.
\ee
The equatorial horizon radius is
\be               \label{A18}
R_0=\left.\sqrt{g_{\varphi\varphi}}\right|_{x=x_h,y=0}=\frac{2}{p}=2M\,,
\ee
while the horizon angular velocity and the linear velocity are
\be
\W_0=\left.\W\right|_{x=x_h}=\frac{pq}{2(1+\sqrt{1-q^2} ) },~~~~
v_0=R_0\W_0=\frac{q}{1+\sqrt{1-q^2} }.
\ee
The polar horizon radius is defined as follows (compare with \eqref{Rp}):
\be          \label{Rpa}
 R_p&=&\frac{2}{\pi} \int_0^1 \left.\sqrt{g_{yy}}\right|_{x=x_h}\,dy~~~\Rightarrow~~~
\frac{R_p}{R_0}=\frac{2}{\pi (1+v_0^2)} \int_0^{\pi/2}\sqrt{1+v_0^2 \cos^2\vartheta }\, d\vartheta.
\ee
Therefore,
one has $R_p/R_0=1$ for $v_0=0$ and $R_p/R_0=0.608$ for $v_0=1$.
 
 The average horizon radius is determined by  the event horizon area $A$ (compare with \eqref{RA}):
 \be                \label{RAa}
 R_A^2=\frac{A}{4\pi}=\int_0^1 \left.\sqrt{g_{yy}g_{\varphi\varphi}}\right|_{x=x_h}\, dy~~~
 \Rightarrow~~~\frac{R_A}{R_0}=\frac{1}{\sqrt{1+v_0^2}}\,.
 \ee
The extremal limit corresponds to $q=1$, hence $\nu=0$ and $x_h=0$.

 \subsubsection{Extremal Kerr solution and its near horizon limit}

Setting $q=1$, $a=M$, and $\nu=0$ in \eqref{prep}--\eqref{Kmtr},  yields the extremal Kerr geometry,
 \be            \label{B16}
 ds^2=-e^{2\V}dt^2+[(x+M)^2+M^2y^2]\left(\frac{dx^2}{x^2}
 +\frac{dy^2}{1-y^2} \right)+x^2(1-y^2) e^{-2\V}\left(d\varphi-\W dt \right)^2,~~~~~~
 \ee
 where
 \be               \label{Kerr33}
e^{2\V}&=&\frac{p^2x^2\,[(px+1)^2+y^2 ] }{\cal Q }, ~~~~
\W=\frac{2p\,(px+1) }{\cal Q },~~~~\nn \\
{\cal Q}&=&4+px\,[8+px\,(7+px\,(px+4)+y^2 ) ] , ~~~~~~p=\frac{1}{M}.
\ee
Let us pass to the rotating frame by setting in \eqref{B16}
\be                        \label{B17}
\varphi=\tilde{\varphi} +\frac{1}{2 M}\, t,
\ee
then rescale the temporal and radial coordinates according to
\be
t=\frac{\tilde{t}}{\epsilon},~~~~~~~~x=\epsilon\, \tilde{x}\,,
\ee
and take the $\epsilon\to 0$ limit. Omitting the tilde signs, the geometry becomes
\be                  \label{nexK}
ds^2=-\frac{x^2\,(1+y^2)}{4 M^2}\, dt^2+M^2(1+y^2)\left(\frac{dx^2}{x^2}+\frac{dy^2}{1-y^2} \right)
+\frac{4 M^2\,(1-y^2)}{(1+y^2)}\left( d\varphi+\frac{x}{2M^2}\, dt\right)^2. ~~~~~~~~
\ee
Since $x=r-M$ describes the deviation from the extremal Kerr horizon, the $\epsilon\to 0$ limit corresponds to
approaching the horizon. Therefore, \eqref{nexK} describes the near-horizon limit of the 
extremal Kerr black hole \cite{Bardeen:1999px}. This is  another, non-asymptotically flat  solution of the vacuum Einstein equations. 
One reads off from \eqref{nexK} the values
\be                \label{B20}
e^{2\V}=\frac{x^2\,(1+y^2)}{4 M^2},~~~~\W=-\frac{x}{2 M^2},
\ee
and these fulfill the Ernst equations \eqref{UUU} with $a^2=0$.
The same expressions can also be obtained directly from \eqref{Kerr33} by 
noting that the transformation \eqref{B17}
implies the shift $\W\to \W-p/2$ (see Eq.\eqref{twis}), after which one sets
$e^{2\V} dt^2=e^{2\tilde{\V}} d\tilde{t}^2$ and $\W dt=\tilde{\W} d\tilde{t}$,
then takes the $\epsilon\to 0$ limit and omits the tilde signs.

 \subsubsection{Embedding diagrams \cite{Smarr:1973zz}} 
 
   \begin{figure}[th]
\hbox to \linewidth{ \hss

			\includegraphics[width=8 cm]{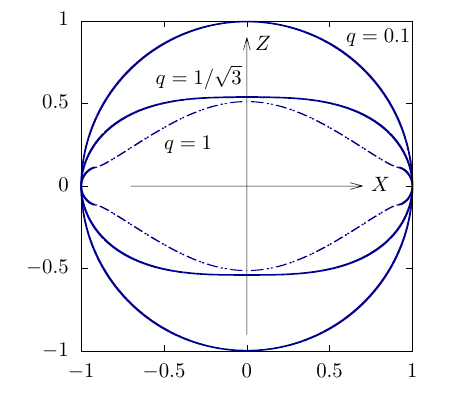}	
	\includegraphics[width=8 cm]{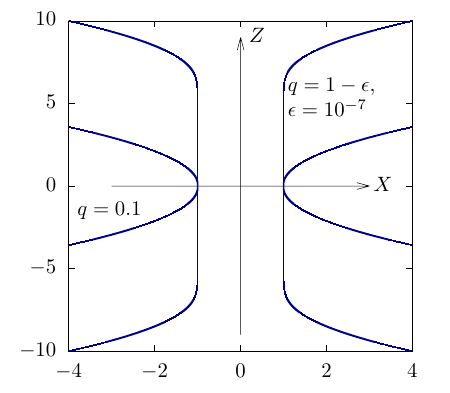}
	
\hss}
\caption{The isometric realization of slices of the $R_0=1$ Kerr geometry for various  $q=a/M$
on surfaces obtained by rotating the $X,Z$ curves
around the $Z$-axis. 
Left: embeddings of the event horizon, with the dashed curve corresponding 
to regions of negative Gaussian curvature.
Right: embedding of the equatorial section; in the extremal limit $q\to 1$ the 
surface approaches the cylinder $X^2+Y^2=1$.}
\label{Fig10}
\end{figure}

 The two-dimensional sections of the Kerr geometry -- the event horizon and the equatorial plane 
 -- are obtained by setting in \eqref{Kmtr}
$t=\mathrm{const}$ and, respectively, either $x=x_h$ or $y=0$. The resulting 2-metric,
together with the corresponding Gaussian curvature, can be represented as
\be
ds^2=A^2(\zeta) d\zeta^2+{\cal R}^2(\zeta) d\varphi^2,~~~~~~~~
 K_{\rm G}=-\frac{1}{A{\cal R}}\left(\frac{{\cal R}^\prime}{A}  \right)^\prime\,.
 \ee
At the horizon one has $\zeta=y\in[-1,1]$, with $A^2=g_{yy}(y)$ and ${\cal R}^2=g_{\varphi\varphi}(y)$
determined by \eqref{Kcomp} at $x=x_{h}$. In the equatorial plane one has $\zeta=x\in[x_h,\infty)$, with
$A^2=g_{xx}(x)$ and ${\cal R}^2=g_{\varphi\varphi}(x)$
computed at $y=0$.

 The same geometry can be obtained by setting
  \be               \label{A24}
 ds^2=dX^2+dY^2\pm dZ^2~\text{with}~
 X+iY={\cal R}(\zeta) e^{i\varphi},~~~Z(\zeta)=\int \sqrt{\pm (A^2- {\cal R}^{\prime 2} ) }\,.
 \ee
This determines a surface of revolution in the $X,Y,Z$ space that is either Euclidean or pseudo-Euclidean.
The $X,Z$ curves corresponding to the $Y=0$ section of this surface are shown in Fig.\ref{Fig10}
for several values of the parameter $q=a/M$. The horizon geometry corresponds to a squashed sphere whose oblateness
increases with $q$. For small $q$ the curvature $K_{\rm G}$ is everywhere positive, and the surface then embeds into
Euclidean space, so one chooses the $``+"$ sign in \eqref{A24}.
For $q=1/\sqrt{3}$ the curvature vanishes at the poles, while
for $q>1/\sqrt{3}$ it becomes negative in the polar regions where $A^2< {\cal R}^{\prime 2}$. 
In these regions one has to
choose the $``-"$ sign in \eqref{A24} to keep the radicand non-negative, 
which corresponds to the dashed curve in Fig.\ref{Fig10}.

 The equatorial section of the Kerr geometry can always be isometrically embedded into Euclidean space.
For small $q$, the corresponding surface is
similar to a paraboloid of revolution,
while for $q\to 1$ it approaches the cylinder $X^2+Y^2=\mathrm{const.}$ 
that characterizes the near-horizon limit of the
extremal Kerr geometry.
Indeed, setting $d\zeta=M\,dx/x$, the equatorial section of the near-horizon geometry \eqref{nexK}
is described by
\be
ds^2=d\zeta^2+R_0^2\, d\varphi^2,
\ee
which is the metric on a cylinder of radius $R_0=2M$.

\renewcommand{\thesection}{Appendix C}
\section{Multipole moments \label{AppC}}

\renewcommand{\theequation}{C.\arabic{equation}}
\setcounter{equation}{0}
\setcounter{subsection}{0}

The systematic definition of multipole moments in General Relativity was
developed by Geroch \cite{Geroch:1970cd} and Hansen \cite{Hansen}.
However, it is worth first recalling the Newtonian theory, 
where the gravity potential $\upphi$ created by a matter source $\mu$ fulfills
$\Delta\upphi=4\pi \mu$. In the axially symmetric case one has, away from the source,
\be               \label{Phi}
\upphi(\vec{R})=-\int \frac{\mu(\vec{r})\, d^3r}{|\vec{R}-\vec{r}\,|} 
=-\sum_{l=0}^\infty \frac{C_l}{R^{l+1}} P_l(\cos\vartheta)
=
-\frac{M}{R}
-\frac{D\,(3\cos^2\vartheta-1)}{2R^3}+\ldots ,
\ee
where the multipole coefficients are determined by the source,
\be           \label{A17}
C_0\equiv  M=\int  \mu (\vec{r})\, d^3 r,~~~~~~~
C_2\equiv D=\frac12\int r^2 \, (3\cos^2\vartheta-1)\,\mu(\vec {r})\,  d^3 r,~~~~\text{etc.}
\ee
The monopole coefficient $C_0$ is the mass,
while the dipole coefficient $C_1$ vanishes if the coordinate origin is chosen at the centre of mass.
The coefficient $C_2=D$ determines the quadrupole moment, which is usually defined as
$Q=D$ (or $Q=2D$ \cite{landau1975}). The quantity $D$
is negative for an oblate matter distribution squeezed toward the equatorial plane and positive for a prolate
distribution elongated along the $z$-axis. For example, this implies that the electric quadrupole moment of prolate atomic 
nuclei  is positive.
However, in General Relativity the quadrupole moment is instead conventionally defined to be positive for
oblate configurations such as spinning stars or black holes, hence one sets
\be       \label{QD}
Q=-D.
\ee

Within the relativistic theory, we have the
field amplitudes $\V,\W,\chioo$ depending on $x,y=\cos\vartheta$ that fulfill
Eqs.\eqref{UUU},\eqref{UUU1}.
One has at large $x$:
\be                         \label{Vwc}
\V&=&-\frac{M}{x}-\frac{D\,(3y^2-1)}{2x^3}+\frac{M\nuu }{3x^3}
+{\cal O}\left(\frac{1}{x^5} \right),\nn \\
\W&=&\frac{2J}{x^3}-\frac{6JM }{x^4}+{\cal O}\left(\frac{1}{x^5} \right), \nn \\
\chioo&=&2Jy\,(3-y^2)+J\left(2J^2-5M^4 \right)\times \frac{ y\,(y^2-1)^2 }{2M^2 x^2}+{\cal O}\left(\frac{1}{x^3} \right),
\ee
where $M,J,D$ are integration constants.
The expression for $\V$ is similar to that for $\upphi$ in \eqref{Phi}, but contains an additional term of order $1/x^3$.
Using the $\V$-equation in \eqref{UUU},
\be
\label{UUUo}
&&[(x^2+\nuu )\V_{,x}]_{,x}+[(1-y^2)\V_{,y}]_{,y}=\frac12\,\rho^2 e^{-4\V}\langle \W,\W\rangle\equiv {\cal S},
\ee
and the boundary conditions in \eqref{bc}, it is not difficult to show that
the parameters $M$ and $D$ can be expressed as integrals of the source,
\be             \label{Mo}
M=\int_{\cal D}  {\cal S} \, dx\, dy,~~~~~~~
D=\frac12\int_{\cal D} \,(2z^2-\rho^2)\,{\cal S} \, dx\,dy+\frac{M\nuu }{3}\,.
\ee
Here the integration is performed over the domain
${\cal D}=\left\{0\leq x\leq \infty,~0\leq y\leq 1\right\}$, while $\rho,z$ are given by \eqref{rz}.
Eq.\eqref{Vwc} implies that
\be                  \label{J21}
J=\left.\frac14\, \chioo\right|_{y=1},
\ee
so that $J$ is determined by the boundary value of $\chioo$ on the symmetry axis, where $y=1$.
This value does not depend on $x$ since, according to Eq.\eqref{twisto}, one has
\be
\partial_x\chioo=\frac{(1-y^2)^2}{x^2+\nuu }\,e^{-4\V}\partial_y w\,~~~~\Rightarrow~~~~\partial_x\chioo(x,y=1)=0,
\ee
therefore
\be
\chioo(x,y=1)=\chioo(x=\infty,y=1)=4J,
\ee
in agreement with the boundary condition in \eqref{bc}.

The parameters $M$ and $J$ can be identified, respectively, with the mass and angular momentum.
Indeed, inserting \eqref{Vwc} into the expressions \eqref{UV} for the
metric coefficients yields
\be
-g_{00}=e^{2\V}-\rho^2\W^2 e^{-2\V}&=&1-\frac{2M}{x}+{\cal O}\left(\frac{1}{x^3} \right),\nn \\
-g_{0\varphi}=\rho^2\W e^{-2\V}&=&\frac{2J\sin^2\vartheta}{x}+{\cal O}\left(\frac{1}{x^2} \right), \nn \\
g_{\varphi\varphi}=\rho^2 e^{-2\V}&=&\left(x+{\cal O}\left(1 \right) \right)^2\sin^2\vartheta.
\ee
The last of these relations shows that near infinity $x$ coincides, to leading order, 
with the Schwarzschild radial coordinate.
Therefore, the other two relations imply that $M$ and $J$ are the mass and angular momentum.

The coefficient $D$ in \eqref{Vwc} is related to the quadrupole moment, 
but the relation is more subtle than in \eqref{QD}.
The procedure for computing multipole moments in General Relativity, 
developed by Geroch \cite{Geroch:1970cd}
and Hansen \cite{Hansen}, is somewhat involved, but it can be simplified by using the
Ernst potentials \cite{Fodor,Fodor1}.

The starting point is the Ernst potential
constructed from the norm and twist of the timelike Killing vector,
which is related to the complex function $\xi$ satisfying \eqref{Er}:
\be            \label{Erpoto}
{\cal E}_{(t)}=e^{2U}+i\chio=  \frac{\xi-1}{\xi+1}~~~~~\Rightarrow~~~~~\xi=\frac{1+{\cal E}}{1-{\cal E~}}.
\ee
The next step is to express $\xi$ in the Weyl coordinates $\rho,z$, and then consider
the expansion of $1/\xi$ on the symmetry axis, where $\rho=0$,
in the $z\to \infty$ limit. Since $\rho=\sqrt{(x^2+\nu )(1-y^2)}$ and $z=xy$, this amounts
to setting $y=1$ and considering
the $x\to\infty$ limit, which yields
\be           \label{neg}
\frac{1}{\xi(x,y=1)}=\sum_{n=0}^\infty \frac{m_{2n}}{x^{2n+1}}+i\sum_{n=0}^\infty \frac{m_{2n+1}}{x^{2n+2}}.
\ee
The multipole moments are determined by the coefficients $m_k$.

For the Kerr solution one has (see \eqref{Kerrxi}) $\xi=px-iqy$, so that
\be
\frac{1}{\xi(x,y=1)}=\frac{1}{px-iq}=\frac{M}{x-ia}=\frac{M}{x}\times \left(1-\frac{ia}{x} \right)^{-1}
=\frac{M}{x}\times  \sum_{k=0}^\infty \frac{(ia)^k}{x^{k}},
\ee
where we used \eqref{rr} to set
\be
\frac{1}{p}=\frac{\rm M}{\MU}\equiv M,~~~~~\frac{q}{p}=\frac{\rm a}{\MU}\equiv a= \frac{J}{M}.
\ee
Therefore, comparing with \eqref{neg}, one obtains the Kerr multipole moments,
\be
m_{2n}=(-1)^{n} M\,a^{2n},~~~~~m_{2n+1}=(-1)^n M\, a^{2n+1},
\ee
which were computed by Hansen \cite{Hansen} (see also \cite{Sotiriou}).
Hansen used the sign convention in which the Kerr quadrupole moment is positive, hence
\be                 \label{Hans}
M=m_0,~~~~~J=m_1=Ma,~~~~~~Q=-m_2=\frac{J^2}{M}=Ma^2.
\ee

Let us now consider more general solutions. The behaviour of $\V,\W,\chioo$ near infinity is given by Eq.\eqref{Vwc},
but at present we need instead a similar asymptotic expansion
for the amplitudes $U,w,\chio$ used in the $t$-parametrization. They are determined by Eqs.\eqref{Ua} and \eqref{www},
\be                 \label{Uwc}
U&=&\V
+{\cal O}\left(\frac{1}{x^4} \right),\nn \\
w&=&\frac{2J\, (y^2-1)}{x}+\frac{2JM (y^2-1) }{x^2}+{\cal O}\left(\frac{1}{x^3} \right), \nn \\
\chio&=&-\frac{2J\, y}{x^2}+\frac{4MJ y }{x^3}+{\cal O}\left(\frac{1}{x^4} \right),
\ee
where the function $\V$ and the integration constants
$M,J,D$ are the same as in Eq.\eqref{Vwc}.
These amplitudes are related to $\V,\W$ in \eqref{Vwc}
via \eqref{UV}.

Computing ${\cal E}_{(t)}=e^{2U}+i\chio$ and using \eqref{Erpoto} to determine $\xi$ yields
\be
\frac{1}{\xi(x,y=1)}=
\frac{M}{x}+\frac{iJ}{x^2}+\frac{3D-M^3-M\nuu }{3x^3}+{\cal O}\left(\frac{1}{x^4} \right)
\equiv \frac{m_0}{x}+i\frac{m_1}{x^2}+\frac{m_2}{x^3}+\ldots
\ee
As a result, $M$ and $J$ are indeed the mass and angular momentum, while the quadrupole moment is
\be             \label{QQQ}
Q&=&-m_2=-D+\frac{M^3+\nuu  M}{3} \nn \\
&=&-\frac14\int_{\cal D} \,(2z^2-\rho^2) e^{-4\V}\langle \W,\W\rangle \, dx\,dy+\frac{M^3}{3},
\ee
where we used \eqref{Mo}.

%%%%%%%%%%%%%%%%%%%%%%%%%

%\bibliographystyle{JHEPmodplain}
%\bibliography{/Users/mvolkov/Dropbox/Horndeski/Bianchi/W,/Users/mvolkov/Dropbox/Horndeski/Bianchi/Cosm}

\providecommand{\href}[2]{#2}\begingroup\raggedright\endgroup

\end{document}